\documentclass[sigconf]{acmart}
\usepackage{multirow}
\usepackage{multicol}
\usepackage{caption}
\usepackage{xcolor} 
\usepackage{color} 
\usepackage{verbatim} 
\usepackage{array}
\usepackage{makecell}
\usepackage{bbding}

\AtBeginDocument{%
  \providecommand\BibTeX{{%
    \normalfont B\kern-0.5em{\scshape i\kern-0.25em b}\kern-0.8em\TeX}}}

\renewcommand\footnotetextcopyrightpermission[1]{}
\setcopyright{acmlicensed}
\copyrightyear{2025}
\acmYear{2025}
\acmDOI{XXXXXXX.XXXXXXX}

\author{Yi Liu\textsuperscript{1}, Xinyi Liu\textsuperscript{1}, Yi Wan\textsuperscript{1}, Panwang Xia\textsuperscript{1}, Qiong Wu\textsuperscript{1},  Yongjun Zhang\textsuperscript{1}\\ \small \textsuperscript{1}School of Remote Sensing and Information Engineering, Wuhan University, Hubei, China\\ \small }

\begin{document}

\title{StereoINR: Cross-View Geometry Consistent Stereo Super Resolution with Implicit Neural Representation}

\begin{abstract}
Stereo image super-resolution (SSR) aims to enhance high-resolution details by leveraging information from stereo image pairs. However, existing stereo super-resolution (SSR) upsampling methods (e.g., pixel shuffle) often overlook cross-view geometric consistency and are limited to fixed-scale upsampling. The key issue is that previous upsampling methods use convolution to independently process deep features of different views, lacking cross-view and non-local information perception, making it difficult to select beneficial information from multi-view scenes adaptively. In this work, we propose Stereo Implicit Neural Representation (StereoINR), which innovatively models stereo image pairs as continuous implicit representations. This continuous representation breaks through the scale limitations, providing a unified solution for arbitrary-scale stereo super-resolution reconstruction of left-right views. Furthermore, by incorporating spatial warping and cross-attention mechanisms, StereoINR enables effective cross-view information fusion and achieves significant improvements in pixel-level geometric consistency. Extensive experiments across multiple datasets show that StereoINR outperforms out-of-training-distribution scale upsampling and matches state-of-the-art SSR methods within training-distribution scales.
\end{abstract}




\begin{teaserfigure}
  \includegraphics[width=\textwidth]{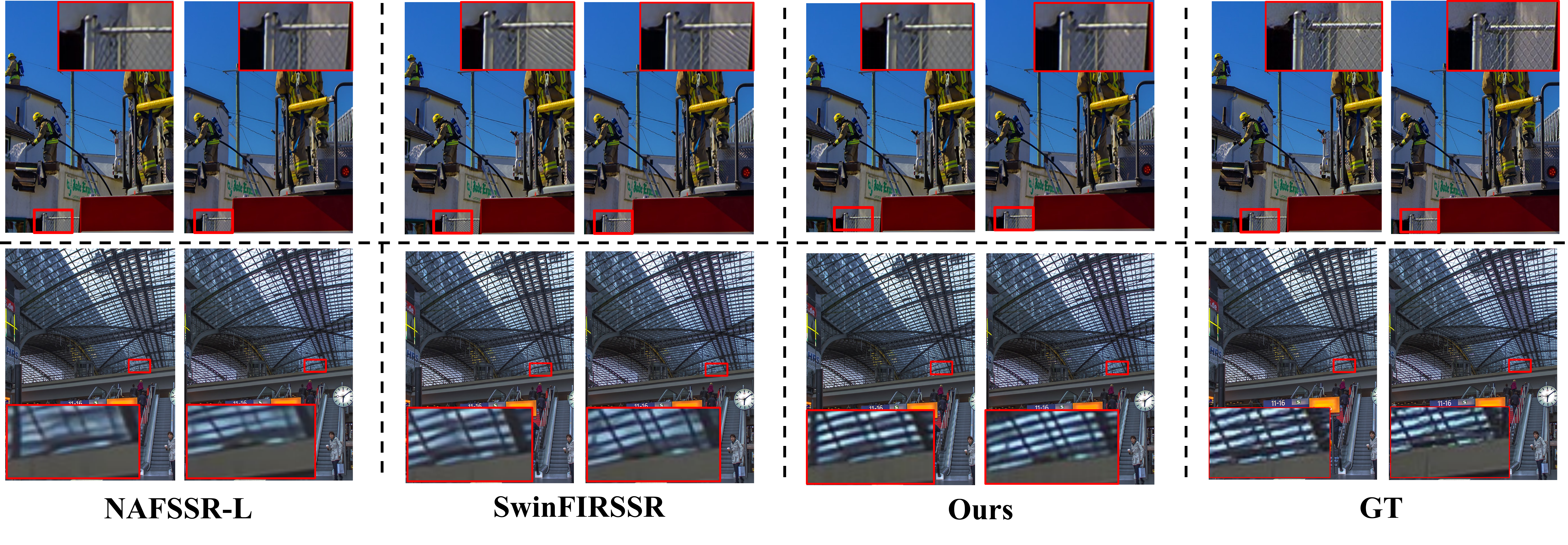}
  \caption{Visual results achieved by NAFSSR\cite{chuNAFSSRStereoImage2022}, SwinFIRSSR\cite{zhangSwinFIRRevisitingSwinIR2023a}, and our method on the Flickr1024\cite{wang2019flickr1024} dataset.}
  \Description{Figure 1}
  \label{fig:fig1}
\end{teaserfigure}

\maketitle

\section{Introduction}
Stereo vision systems are crucial for a range of applications, including 3D reconstruction and autonomous navigation, where high-resolution images are essential for accurate depth estimation and scene understanding. However, existing methods rely on fixed-scale upsampling, which introduces two key issues: 1) These predefined fixed-scale upsampling methods are inflexible. Once the upsampling factor is set, it cannot be adjusted without modifying the model architecture. This limitation severely restricts their practical use in real-world systems that demand adaptable resolution adjustments; 2) Fixed-scale upsampling methods process the local information of each view independently, lacking guidance from global and cross-view information. Therefore, it is urgent to develop novel stereo-upsampling methods that can generate geometrically consistent structures across views based on super-resolution at arbitrary scales.

\begin{figure}[t]
\centering
\vspace{4mm}
\includegraphics[width=0.96  \linewidth]{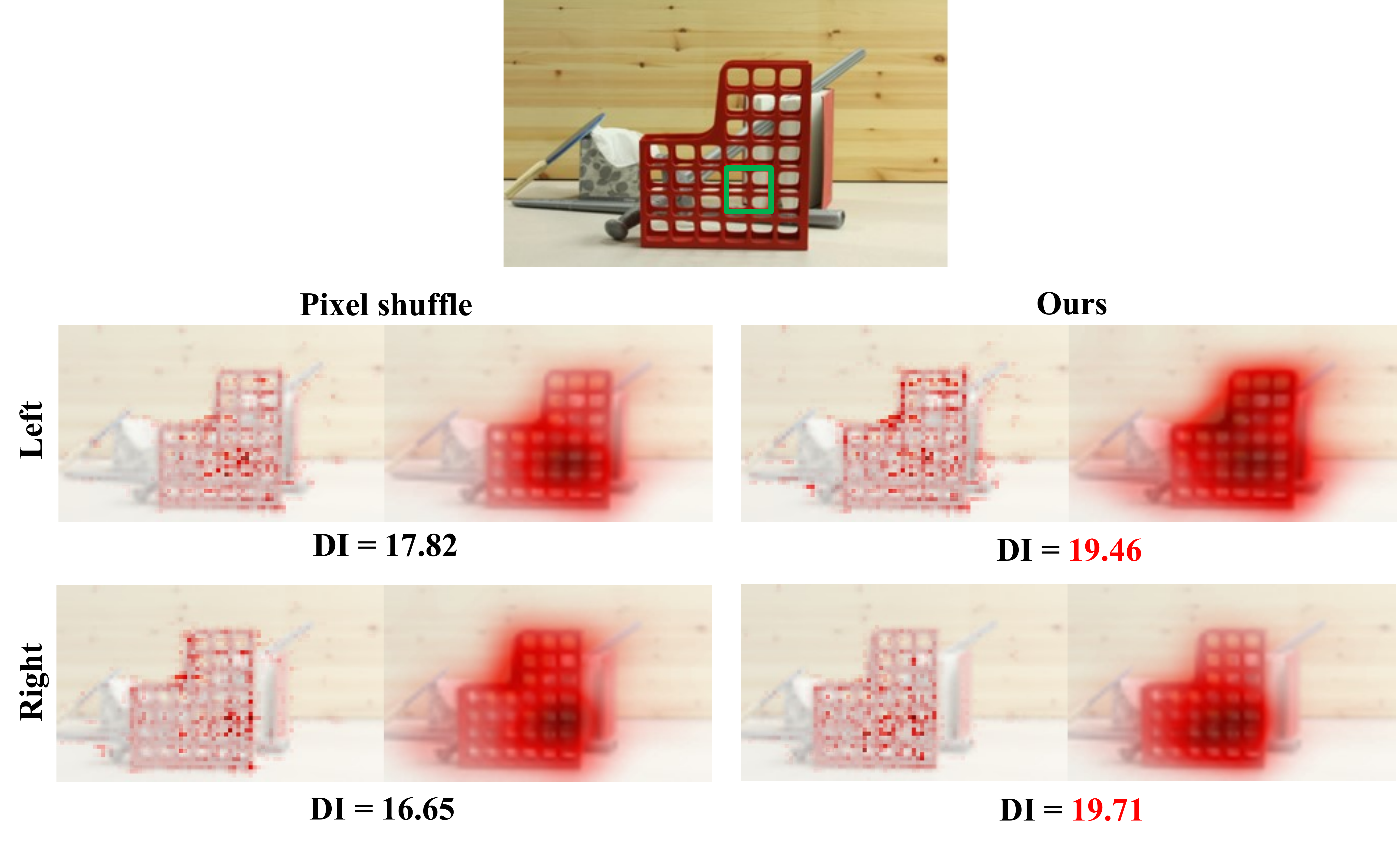}
\caption{Local attribution map (LAM) results achieved by different upsampling methods on the Flickr1024 \cite{wang2019flickr1024} dataset. The first row shows the super-resolution results for the left view, while the second and third rows display the LAM results for the left and right views, respectively. Red pixels indicate that the pixel participates in the reconstruction of the pixels in the green rectangle in the result of the left view. DI (Diffusion Index) indicates the range of involved pixels, with a higher DI indicating a wider receptive field.}
\label{fig.lam_for_upsampler}
\end{figure}

Arbitrary-scale stereo super-resolution (ASSSR) aims to employ a single model to upscale low-resolution (LR) image pairs to high-resolution (HR) outputs at any appropriate real-valued magnification. Existing stereo super-resolution (SSR) methods ~\cite{chuNAFSSRStereoImage2022,dai2021feedback,maPerceptionOrientedStereoImage2022,songStereoscopicImageSuperResolution2020,wang2019learning,wangParallaxAttentionUnsupervised2022,yaoCrossAttentionCoupledUnmixing2020,zhangStereoImageRestoration2024,zhangSwinFIRRevisitingSwinIR2023a,zouCrossViewHierarchyNetwork2023,qiuSCNAFSSRPerceptualOrientedStereo2023} typically rely on fixed-scale upsampling techniques (e.g., transposed convolution, pixel shuffling), which limit their flexibility in real scenarios where resolution requirements may vary. Therefore, we introduce Implicit Neural Representation (INR) ~\cite{chenLearningContinuousImage2021} for arbitrary-scale upsampling, which has been widely used in single image super-resolution~\cite{xuUltraSRSpatialEncoding2022,lugmayrSRFlowLearningSuperResolution2020, songOPESROrthogonalPosition2023,yangImplicitTransformerNetwork, lee2022local,chen2023cascaded, caoCiaoSRContinuousImplicit2023, Wei_2023_CVPR} and video super-resolution~\cite{chenVideoINRLearningVideo2022, shang2024arbitrary}. INR treats images as continuous spatial representations and achieves arbitrary-scale super-resolution by learning a coordinate-to-color mapping. Compared to the pixel shuffle method, which integrates local information through convolution, the advantage of the INR method lies in its ability to incorporate attention mechanisms to capture nonlocal and cross-view information. As demonstrated in Figure \ref{fig.lam_for_upsampler}, with INR, our method is able to leverage more pixel information from both views to reconstruct the high-resolution results. Motivated by this, we replace the convolution-based upsampling methods with INR to enhance the cross-view consistency of stereo super-resolution results. This approach not only has the potential to promote geometric consistency in super-resolved images but also overcomes the limitations of fixed-scale methods, offering an effective solution for arbitrary-scale stereo super-resolution.

The performance of ASSSR is intricately tied to the representation capacity of the encoded latent features. However, due to the challenge of acquiring paired stereo image data, available stereo training datasets are often limited in scale and quality, leading to suboptimal reconstruction of texture details. In contrast, single-image datasets are more abundant and generally of higher quality. Single-image super-resolution (SISR) methods such as IPT\cite{chen2021pre} and HAT\cite{chen2023activating} leverage large-scale datasets like ImageNet\cite{deng2009imagenet} for pre-training to explore the potential of Transformer. Moreover, ASteISR\cite{zhou2024asteisr} has demonstrated the feasibility of adapting SISR models to the stereo image domain. Inspired by these methods, we fine-tune the state-of-the-art SISR model on the stereo dataset to fully leverage the prior knowledge.

In this paper, our goal is to achieve arbitrary-scale stereo image super-resolution and accurately reconstruct geometrically consistent and coherent details across different viewpoints, ensuring structural integrity and realism of high-resolution images in multi-view scenarios. To this end, we proposed StereoINR, which combines two basic components: 1) latent code encoder, and 2) disparity-guided arbitrary-scale up-sampler (DGASU). The latent code encoder captures cross-view correlations, extracts deep features from the left and right images, and provides rich feature representations for subsequent up-sampling reconstruction. The disparity-guided arbitrary-scale upsampling module is based on INR and uses the disparity between the left and right images as guidance. It employs cross-attention to extract beneficial spatial information from different viewpoints, optimizing the upsampling results while maintaining multi-view consistency. This approach helps to maintain the consistency of the cross-view geometry and extract subpixel details from different viewpoints. Besides, since the input coordinates are continuous, we can query image values at any scale, enabling arbitrary scale up-sampling. We highlight our main contributions as follows:
\begin{itemize}
    \item We propose a novel stereo super-resolution framework, named StereoINR. During the upsampling stage, we utilize INR to leverage non-local and cross-view information instead of independently processing features from different viewpoints, enhancing the consistency between the left and right views.
    \item By learning the coordinate-to-color mapping, our method provides a unified solution for stereo image super-resolution across different scales.
    \item Extensive experiments on multiple public datasets demonstrate the superior performance of our method in achieving high-fidelity stereo image reconstruction. Our framework not only enhances stereo geometry consistency but also effectively handles arbitrary-scale super-resolution, outperforming existing approaches in both qualitative and quantitative evaluations.
\end{itemize}

\begin{figure*}[t]
% \vspace{-1.1cm}
\centering
\includegraphics[width=0.96  \textwidth]{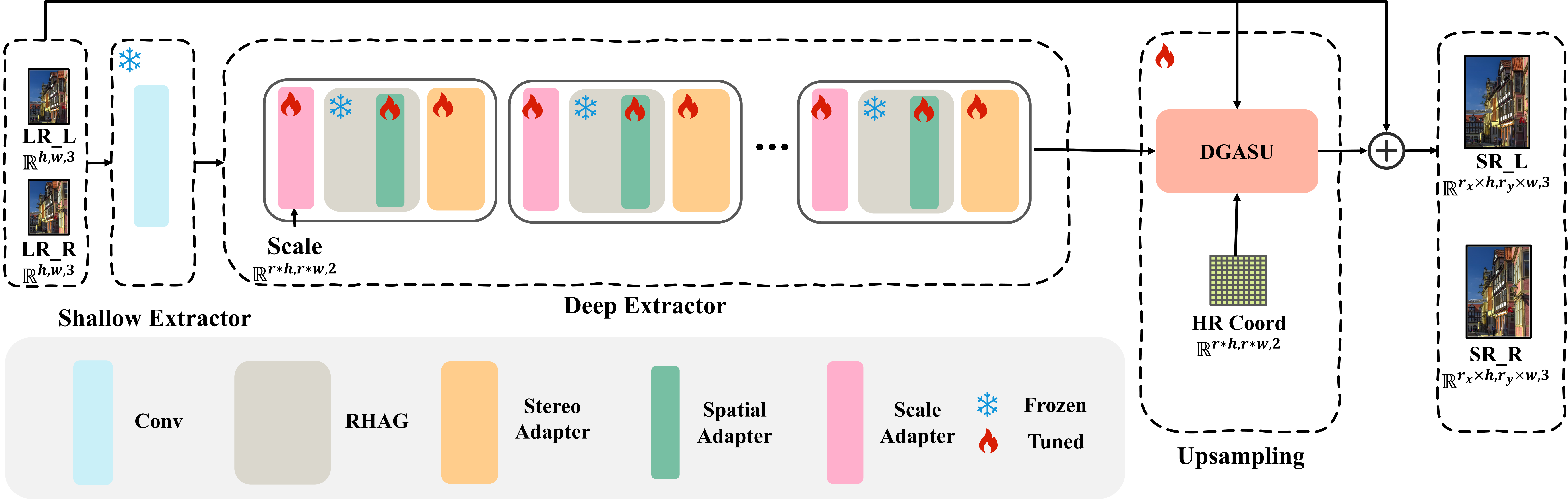}
\vspace{-2mm}
\caption{The overall architecture of proposed method. RHAG is the basic block in HAT\cite{chen2023activating} and is frozen during training. The spatial adapter is embedded in the Feed Forward stage of RHAG. The coordinates are randomly sampled from HR image. }
\vspace{-2mm}
\label{fig.overall}
\end{figure*}

\section{Related Work}
\noindent\textbf{Implicit Neural Representation.}
Implicit Neural Representation was initially applied in 3D vision, where NeRF ~\cite{mildenhall2021nerf} uses spatial coordinates and light direction to implicitly model a 3D scene. Later,  LIIF~\cite{chenLearningContinuousImage2021} introduces implicit neural representations into the field of 2D images, employing MLPs to map image coordinates to RGB values for continuous image representation. Since then,  numerous studies ~\cite{caoCiaoSRContinuousImplicit2023,chenVideoINRLearningVideo2022,lugmayrSRFlowLearningSuperResolution2020,songOPESROrthogonalPosition2023,xuUltraSRSpatialEncoding2022,yangImplicitTransformerNetwork} have leveraged improved implicit neural representations instead of traditional up-sampling methods such as pixel shuffling. The merit lies in the fact that INR considers images as continuous representations rather than discrete values, allowing the mapping of continuous 2D coordinates to pixel values and enabling flexible up-sampling at any scale. Therefore, image implicit neural representations have shown remarkable potential in image restoration ~\cite{chen2024bidirectional}, semantic segmentation~\cite{sarkar2023parameter,gong2023continuous}, optical flow ~\cite{jungAnyFlowArbitraryScale2023}, disparity estimation ~\cite{liangAnyStereoArbitraryScale2024}, and so on. 

\begin{figure}[t]
\centering
\includegraphics[width=0.96  \linewidth]{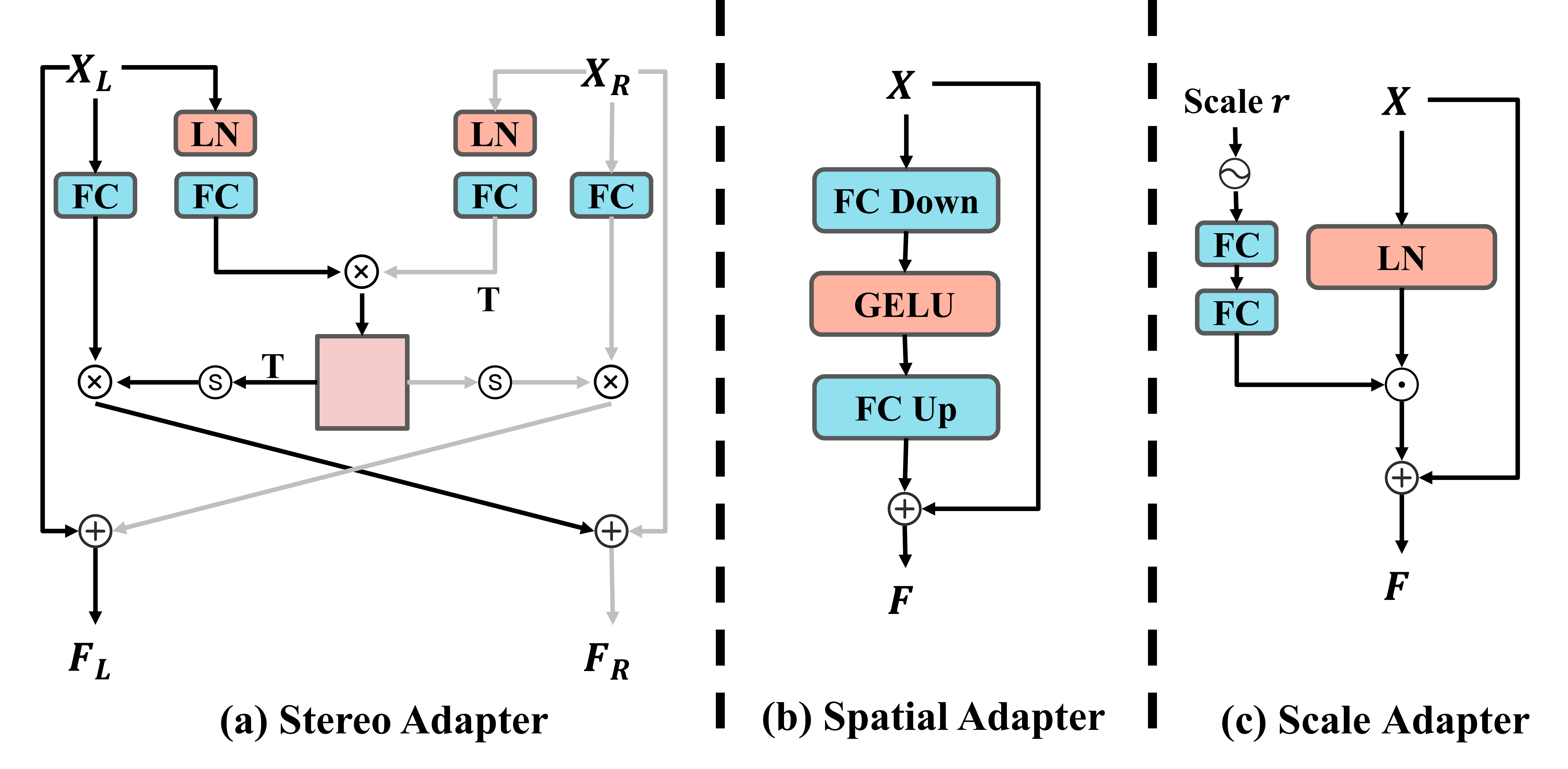}
\vspace{-2mm}
\caption{The architecture of the adapter. (a) Stereo adapter, which uses stereo cross attention to achieve left and right view information interaction; (b) Spatial adapter, which modulates input features to adapt to stereo super-resolution; (c) Scale adapter, which uses scale information to modulate input features.}
\vspace{-6mm}
\label{fig.Adapter}
\end{figure}

\noindent\textbf{Stereo Super Resolution.}
Stereo Super Resolution aims to recover high resolution details from a pair of binocular low-resolution images. StereoSR~\cite{jeon2018enhancing} proposes to leverage stereo images to enhance spatial resolution, which jointly trains a two-stage network of luminance and chrominance to learn the mapping of low and high resolution images. Since the epipolar constraint ~\cite{wangParallaxAttentionUnsupervised2022} of the stereo images, we need to pay more attention to the correlation between pixels on the same row on the left and right images. PASSRNet~\cite{wang2019learning} proposed a parallax attention module to calculate the correlation volume of pixels along the epipolar line. Recently, many methods~\cite{chuNAFSSRStereoImage2022,zhangStereoImageRestoration2024,yaoCrossAttentionCoupledUnmixing2020,ying2020stereo, zhangSwinFIRRevisitingSwinIR2023a} have introduced various cross-attention modules that demonstrate strong performance in stereo super-resolution. However, these approaches typically restrict cross-view information interaction to the deep feature extraction stage, resulting in isolated left and right views during the feature upsampling process. To overcome this limitation, we propose a novel upsampling module that leverages disparity constraints inherent in stereo imaging. This module warps deep features using disparity information and incorporates the warped features as auxiliary data, enabling the recovery of high-resolution geometries that are consistent across views.

\noindent\textbf{Parameter Efficient Fine-tuning in Super Resolution. }
Pre-training \& fine-tuning can enhance the transferring efficiency and performance in many high-level visual tasks. In low-level vision, IPT\cite{chen2021pre} demonstrates the benefits of multi-task training—covering denoising, deraining, and super-resolution, while EDT\cite{li2021efficient} pretrains on specific tasks with varied degradation levels, both achieving remarkable results. These approaches underscore the transferability of pretraining across different tasks. For a single task, pre-training leverages the diverse data distributions available in large-scale datasets.  HAT\cite{chen2023activating} demonstrates this by pretraining its model on ImageNet and subsequently fine-tuning on target datasets, thereby enhancing the representation capability of the Transformer architecture. Moreover, ASteISR \cite{zhou2024asteisr} is the first to propose transferring SISR methods to the field of stereo vision. Inspired by these works, we fine-tune the HAT encoder on the stereo datasets using adapter tuning.

\section{Methods}

\subsection{Overview}
Given any up-sample scale $r\in \mathbb{R}$, we aim to reconstruct high resolution stereo image pairs $I_L^{SR},I_R^{SR}\in \mathbb{R}^{r\times h,r\times w,3}$ from low resolution inputs $I_L^{LR},I_R^{LR} \in \mathbb{R}^{h,w,3}$. Given a 2D query coordinates $x$ in a continuous image domain, the objective of arbitrary-scale stereo super-resolution is to establish a precise mapping between query coordinates and pixel color values:
\begin{equation}
  s_L,s_R = I^{SR}_L(x),I^{SR}_R(x) = \mathcal{F}(\mathcal{E}(I_L^{LR},I_R^{LR}),x; r),
  \label{eq:eq_inr}
\end{equation}
where $s_L,s_R$ represent the predicted RGB pixel values corresponding to the coordinate $x$ in the left and right views, respectively. $\mathcal{E}$ denotes the latent code encoder, and $\mathcal{F}$ denotes the disparity-guided arbitrary-scale up-sampling module. An overview of the proposed framework is presented in Figure \ref{fig.overall}. 

% On the one hand, arbitrary-scale stereo super-resolution methods largely rely on the representation capability of latent codes, which effectively capture and compress complex information in input images, influencing the quality of super-resolution reconstruction. Generally, an encoder with stronger representation capability can extract image features more comprehensively, providing richer details and more accurate pixel reconstruction. On the other hand, existing upsampling methods overlook the geometric structure consistency under stereo imaging conditions and instead process information from different viewpoints in isolation. Motivated by this, we aim to integrate beneficial information from multiple viewpoints during the upsampling stage to enhance the viewpoint consistency of super-resolution reconstruction results. To address the aforementioned challenges, we propose StereoINR, which adapts the powerful representation capability of HAT\cite{chen2023activating} to the domain of stereo super-resolution. Moreover, it utilizes disparity to guide the deep fusion of latent representations from the left and right images, thereby enhancing cross-view consistency in the reconstruction process. 

\subsection{Adapter tuning latent encoder}
Since HAT is designed as a fixed-scale single-image super-resolution method, it lacks awareness of cross-view contextual relationships and multi-scale feature interactions, making its direct transfer to the stereo imaging domain inappropriate. To address this issue, we embed stereo, scale, and spatial adapters into the basic module of HAT, the Residual Hybrid Attention Group (RHAG) module \cite{chen2023activating}. The architecture of these adapters is shown in Figure \ref{fig.Adapter}.

Specifically, the stereo adapter leverage cross-attention to exchange information from left and right views, which is similar to the SCAM in NAFSSR\cite{chuNAFSSRStereoImage2022}. The spatial adapter is embedded within the attention module of RHAG, where it compresses and then restores feature dimensions. It projects the original input dimension $d$ to a lower-dimensional space $m$, allowing control over the number of parameters in the adapter by adjusting the value of $m$. Typically, $m \ll d$. In the output stage, a second feed-forward sub-layer restores the original input dimension, projecting $m$ back to $d$ as the final output of the spatial adapter module. As for the scale adapter, we adopt a structure similar to the channel attention block, modulating scale information along the channel dimension, which can be formulated as:
\begin{align}
  e&=\mathrm{MLP}(s),\\
  x_{out}&=\mathrm{LN}(x_{in})\odot \mathrm{Sigmoid}(\mathrm{MLP}(e))+x_{in}.
  \label{eq:eq_scale_adapter}
\end{align}
where $\odot$ denotes the element-wise multiplication, $\mathrm{MLP}$ denotes the MLP projector, $\mathrm{LN}$ denotes LayerNorm, $x_{in}$ is the input feature, and $x_{out}$ is the feature after modulation.
% Specifically, we introduce a Stereo Adapter and a Scale Adapter into the HAT model and replace its upsampling module with the disparity-guided arbitrary scale upsampler. 

It is noteworthy that our method differs from ASteISR\cite{zhou2024asteisr} in that we aim for making the features extracted by the Encoder universally applicable to detail reconstruction at arbitrary scales. To accomplish this, we explicitly inject scale-specific information into the scale adapter, allowing dynamic parameter modulation that adaptively adjusts features for different scaling factors. This design fundamentally enhances the model's generalization capability for multi-resolution tasks, setting it apart from fixed-scale approaches.

Finally, the latent codes for the left and right viewpoints are obtained through the fine-tuned encoder $\mathcal{E}$.
\begin{equation}
  z_{L}, z_{R} = \mathcal{E}(I_L^{LR},I_R^{LR})
  \label{eq:eq_encoder}
\end{equation}

\begin{figure}[t]
\centering
\includegraphics[width=0.96  \linewidth]{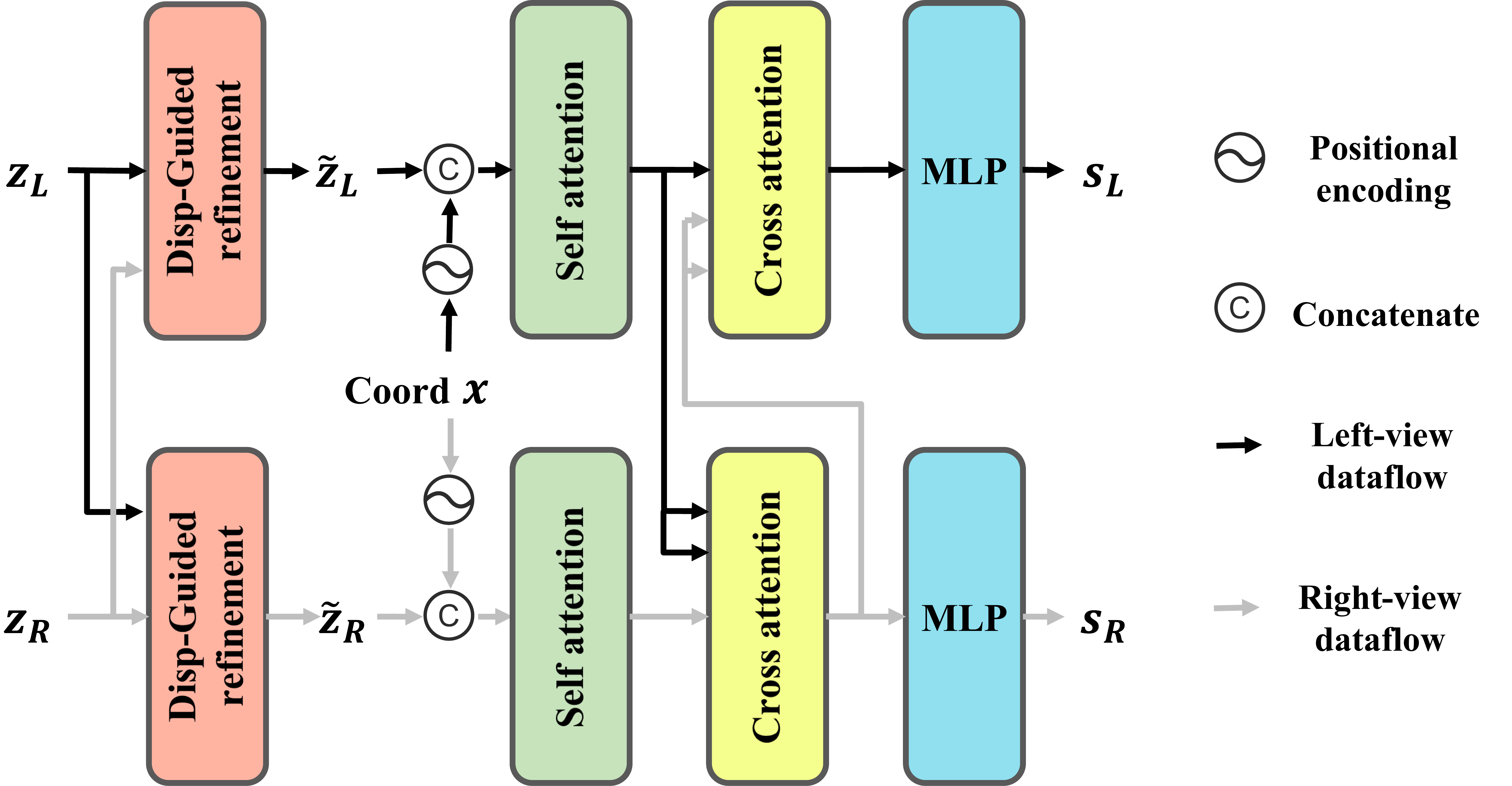}
\vspace{-2mm}
\caption{The overall architecture of DGASU. Here, the self-attention and cross-attention are both the simple dot-product attention.}
\vspace{-4mm}
\label{fig.DGASU}
\end{figure}

\begin{figure}[t]
\centering
\includegraphics[width=0.96  \linewidth]{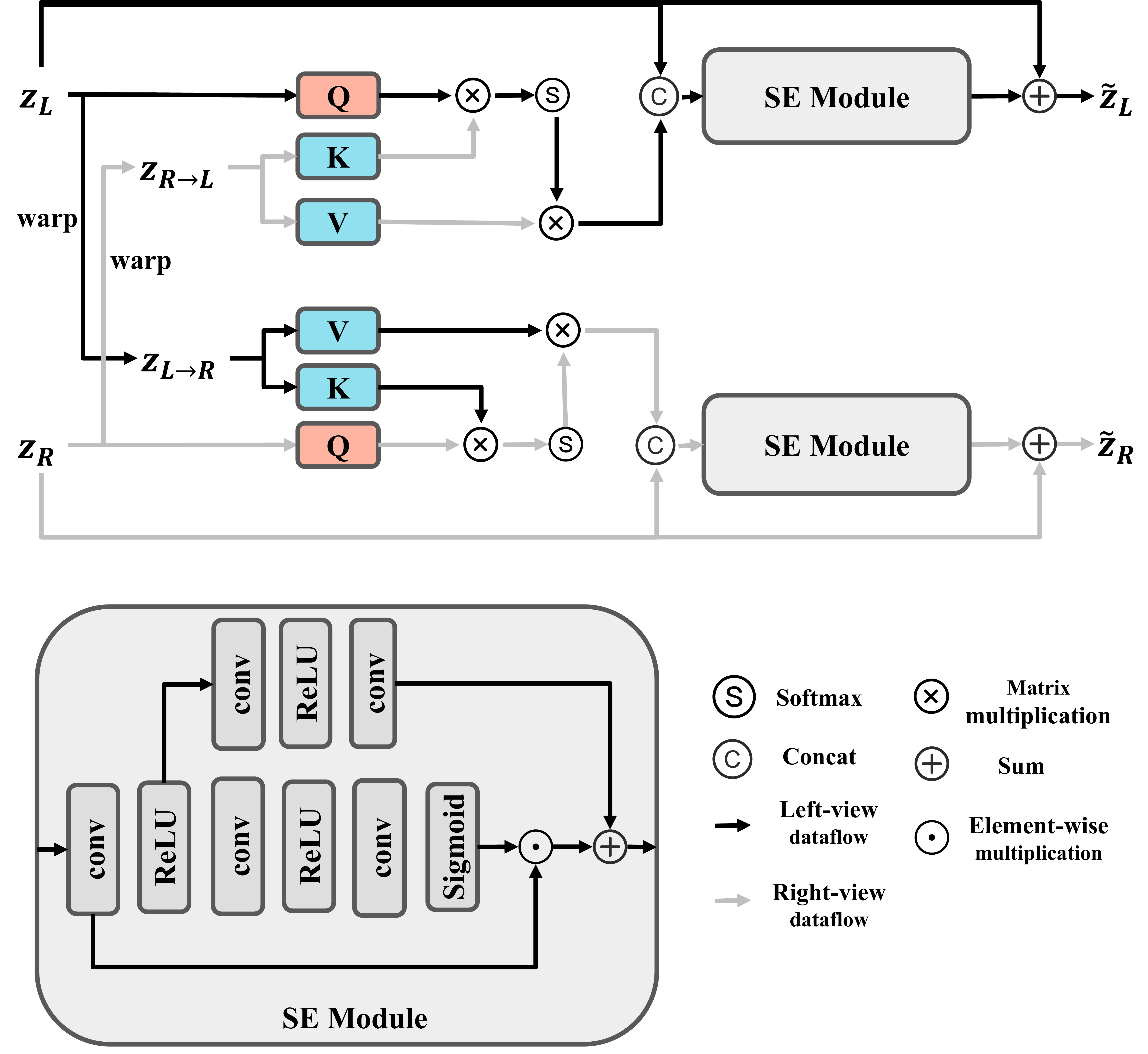}
\vspace{-2mm}
\caption{The overall architecture of disparity-guided refinement module.}
\vspace{-4mm}
\label{fig.DGRF}
\end{figure}

\subsection{Disparity guided arbitrary scale upsampler}
After getting the latent codes with the deep feature extractor, we aim at generating a stereo continuous image representation for stereo image pairs. Initially, we estimate the disparity between the left and right view images:
\begin{align}
  d_{L\rightarrow R} = \mathrm{disp}(I_{L}^{LR},I_{R}^{LR}),\\
  d_{R\rightarrow L} = \mathrm{disp}(I_{R}^{LR},I_{L}^{LR}).
  \label{eq:eq_dispest}
\end{align}
where $\mathrm{disp}(\cdot)$ denotes a state-of-the-art disparity estimator. $d_{L\rightarrow R}$ is employed to align the latent encoding of the right view with the left view, while $d_{R\rightarrow L}$ is used to align the latent encoding of the left view with the right view:
\begin{align}
  z_{L\rightarrow R} = \mathrm{warp}(z_{R}, d_{L\rightarrow R}),\\
  z_{R\rightarrow L} = \mathrm{warp}(z_{L}, d_{R\rightarrow L}).
  \label{eq:eq_warp}
\end{align}
where $\mathrm{warp}(\cdot)$ denotes the standard image/feature warping operation using the bilinear kernel. 

To address the challenge of information extraction under complex multiview conditions while mitigating occlusion interference, we employ a cross-attention mechanism for adaptive spatial information aggregation across viewpoints. This architecture uses attention-based feature alignment to dynamically compute inter-view attention scores, enabling: 1) selective emphasis on geometrically consistent and semantically complementary cues from non-occluded regions, and 2) effective suppression of conflicting signals due to texture ambiguities, parallax discontinuities, and partial occlusions. The learned attention weights facilitate optimal information fusion, enhancing feature discriminability while maintaining robustness against cross-view appearance variations. As depicted in Figure \ref{fig.DGRF}, the query $(Q_{L}, Q_{R})$ is derived from the latent codes of the current viewpoint, while the key $(K_{R\rightarrow L}, K_{L\rightarrow R})$ and the value $(V_{R\rightarrow L}, V_{L\rightarrow R})$ are constructed through aligned latent codes:
\begin{align}
  Q_{L}=W_{q}^{L}z_{L},  K_{R\rightarrow L} = W_{k}^{L}z_{R\rightarrow L}, V_{R\rightarrow L} = W_{v}^{L}z_{R\rightarrow L},\\
  Q_{R}=W_{q}^{R}z_{R},  K_{L\rightarrow R} = W_{k}^{R}z_{L\rightarrow R}, V_{L\rightarrow R} = W_{v}^{R}z_{L\rightarrow R}.
  \label{eq:eq_qkv}
\end{align}

Then we calculate the similarity between queries and keys to generate learnable attention weights, which are applied to perform a weighted summation of value vectors, ultimately achieving cross-view feature fusion:
\begin{align}
  F_{L}=\mathrm{softmax} (\frac{Q_{L}K_{R\rightarrow L}}{\sqrt{C}})V_{R\rightarrow L},\\
  F_{R}=\mathrm{softmax} (\frac{Q_{R}K_{L\rightarrow R}}{\sqrt{C}})V_{L\rightarrow R}.
  \label{eq:eq_crossfusion}
\end{align}
where $C$ denotes the feature channel of the query vectors. To further enhance feature selection and aggregation, we implement a variant of the squeeze and excitation mechanism as a form of global self-attention. It computes the refined features $(\widetilde{z}_{L},\widetilde{z}_{R})$ from $(F_{L},F_{R})$ with reference to $(z_{L},z_{R})$:
\begin{align}
  \widetilde{z}_{L}=z_{L}+\mathrm{Conv}(F_{L})+ F_{L}\odot\mathrm{sigmoid}(\mathrm{Conv}(F_{L})),\\
  \widetilde{z}_{R}=z_{R}+\mathrm{Conv}(F_{R})+ F_{R}\odot\mathrm{sigmoid}(\mathrm{Conv}(F_{R})).
  \label{eq:eq_se}
\end{align}

Then, given a 2d coordinate $x$, we formulate the coordinate-to-color mapping as an MLP $\mathcal{F}_{\theta}:\mathbb{R}^{2+C}\rightarrow\mathbb{R}^3$ that predicts the corresponding RGB values conditioned on spatial coordinates, where $\theta$ denotes the learnable network parameters.
\begin{equation}
  s_{L},s_{R} = \mathcal{F}_{\theta}(\widetilde{z}_{L}^{\ast}, \widetilde{z}_{R}^{\ast}, x-x^{\ast}).
  \label{eq:eq_stereoinr}
\end{equation}
where $(\widetilde{z}_{L}^{\ast}, \widetilde{z}_{R}^{\ast})$ are the nearest latent codes from the query coordinate $x$, $x^{\ast}$ is the coordinate of the latent codes $(\widetilde{z}_{L}^{\ast}, \widetilde{z}_{R}^{\ast})$ in the image domain.

Vanilla MLPs that employ ReLU activation functions exhibit inherent spectral bias~\cite{rahaman2019spectral,lee2022local}, tending to capture low-frequency content. To address this problem, we map the input coordinates into a high-dimensional space by sinusoidal positional encoding and concatenate it with the latent codes.
\begin{align}
  \gamma(x)=\mathrm{concat}(x,cos(\pi x),sin(\pi x),\dots,cos(n\pi x),sin(n\pi x)).
  \label{eq:eq_position_encoding}
\end{align}

Previous INR methods \cite{Wei_2023_CVPR} aggregated features by stacking multiple self-attention layers. We improve feature aggregation by alternating between self-attention and cross-attention, facilitating more effective information exchange between the left and right views. As shown in Figure \ref{fig.DGASU}, we aggregate multiple views information using alternating self-attention and cross-attention mechanisms. Finally, the MLP is adapted to output the predicted color values, which are then combined with the bicubic-upsampled input image pair through a LR skip connection to obtain the final super-resolution results.
\begin{equation}
  s_{L},s_{R} = \mathcal{F}_{\theta}(\widetilde{z}_{L}^{\ast}, \widetilde{z}_{R}^{\ast}, \gamma(x-x^{\ast})))+\mathrm{Bilinear}(I_L^{LR}(x),I_R^{LR}(x)).
  \label{eq:eq_stereoinr}
\end{equation}

\begin{table*}[t]
\centering
\caption{Quantitative comparison with state-of-the-art methods for \underline{\textbf{arbitrary-scale SSR}} on Middlebury~\cite{scharstein2014high} (\textbf{PSNR} $\uparrow$ /\textbf{LPIPS} $\downarrow$). \textcolor{red}{Red} and \textcolor{blue}{blue} colors indicate the best and the second-best performance, respectively.}
\vspace{-2mm}
\footnotesize{
% \begin{tabular}{c|ccc|ccccc}
\resizebox{\textwidth}{!}{
\small
\begin{tabular}{c
|>{\centering\arraybackslash}p{0.10\textwidth}>{\centering\arraybackslash}p{0.10\textwidth}>{\centering\arraybackslash}p{0.10\textwidth}
|>{\centering\arraybackslash}p{0.10\textwidth}>{\centering\arraybackslash}p{0.10\textwidth}
>{\centering\arraybackslash}p{0.10\textwidth}>{\centering\arraybackslash}p{0.10\textwidth}>{\centering\arraybackslash}p{0.10\textwidth}
}
\toprule
\multirow{2}{*}{Method} 
% & \multicolumn{3}{c|}{In-scale} 
& \multicolumn{8}{c}{\textit{PSNR} $\uparrow$ /\textit{LPIPS} $\downarrow$} \\
\cmidrule(lr){2-9} 
 & $\times2$ & $\times3$ & $\times4$ 
  & $\times6$ & $\times12$ & $\times18$ & $\times24$ & $\times30$\\
\midrule\midrule
Bicubic 
& 30.99/0.2101 & 26.80/0.3060 & 25.17/0.3815 
& 22.99/0.4733 & 20.01/0.6140 & 18.32/0.6748 & 17.45/0.6977 & 16.76/0.7129 \\
EDSR-LIIF ~\cite{chenLearningContinuousImage2021} 
& {34.79/0.1042} & {31.18/0.2313} & {29.01/0.2958} 
&{26.75/0.4112} & {23.25/0.5625} & {21.51/0.6525} & {20.08/0.6921} & {19.14/0.7166} \\
EDSR-LTE ~\cite{lee2022local} 
& {34.87/0.1039} & {31.26/0.2291} & {29.11/0.2936}
& {26.79/0.4083} & {23.25/0.5603} & {21.51/0.6482} & {20.15/0.6908} & {19.19/0.7142} \\
EDSR-LINF ~\cite{yao2023local} 
& 34.89/0.1061 & 31.29/0.2121 & 29.13/0.2989 
& 26.77/0.3929 & 23.20/0.5482 & 21.50/0.6227 & 20.05/0.6727 & 19.12/0.6944\\
EDSR-OPESR ~\cite{songOPESROrthogonalPosition2023} 
& 34.74/0.1047 & 31.27/0.2278 & 29.14/0.2919 
& 26.82/0.4082 & 23.31/0.5614 & 21.59/0.6482 & \textcolor{blue}{20.42}/0.6903 & \textcolor{blue}{19.56}/0.7150 \\
\midrule
%%%%%%%%%% Transformer %%%%%%%%%%
SwinIR-LIIF ~\cite{chenLearningContinuousImage2021} 
& {34.96/0.0934} & {31.30/0.2114} & {29.23/0.2746} 
&{26.82/0.3940} & {23.36/0.5417} & {21.47/0.6330} & {20.12/0.6752} & {19.41/0.7057} \\
SwinIR-LTE ~\cite{lee2022local} 
& {35.66/0.0932} & {32.00/0.2109} & {29.78/0.2741}
& {27.29/0.3912} & {23.72/0.5392} & {21.76/0.6310} & {20.36/0.6714} & {19.35/0.7014} \\
SwinIR-LINF ~\cite{yao2023local} 
& 35.61/0.0960 & 31.95/\textcolor{blue}{0.1920} & 29.66/0.2789 
& 27.26/\textcolor{blue}{0.3746} & \textcolor{blue}{23.56/0.5174} & \textcolor{blue}{21.74}/\textcolor{red}{0.5941} & 20.29/\textcolor{blue}{0.6493} & 19.23/0.6786\\
EQSR~\cite{wang2023eqsr} 
& \textcolor{blue}{35.81/0.0907} & \textcolor{blue}{32.20}/{0.2056} & \textcolor{blue}{30.02/0.2652} 
& \textcolor{blue}{27.43}/{0.3875} & {23.52/0.5354} & {21.46/0.6269} & {20.04/0.6542} & {19.19}/\textcolor{blue}{0.6640}\\
\midrule
StereoINR (\textbf{Ours})
& \textcolor{red}{36.43/0.0787} & \textcolor{red}{32.48/0.1807} & \textcolor{red}{30.47/0.2349} 
& \textcolor{red}{27.64/0.3590} & \textcolor{red}{23.98/0.5015} & \textcolor{red}{21.89}/\textcolor{blue}{0.6000} & \textcolor{red}{20.57/0.6372} & \textcolor{red}{19.74/0.6630} \\
\bottomrule
\end{tabular}
}
}
\label{tab:Quan_Middlebury}
\end{table*}

\begin{figure*}[t]
\centering
\includegraphics[width=0.96  \textwidth]{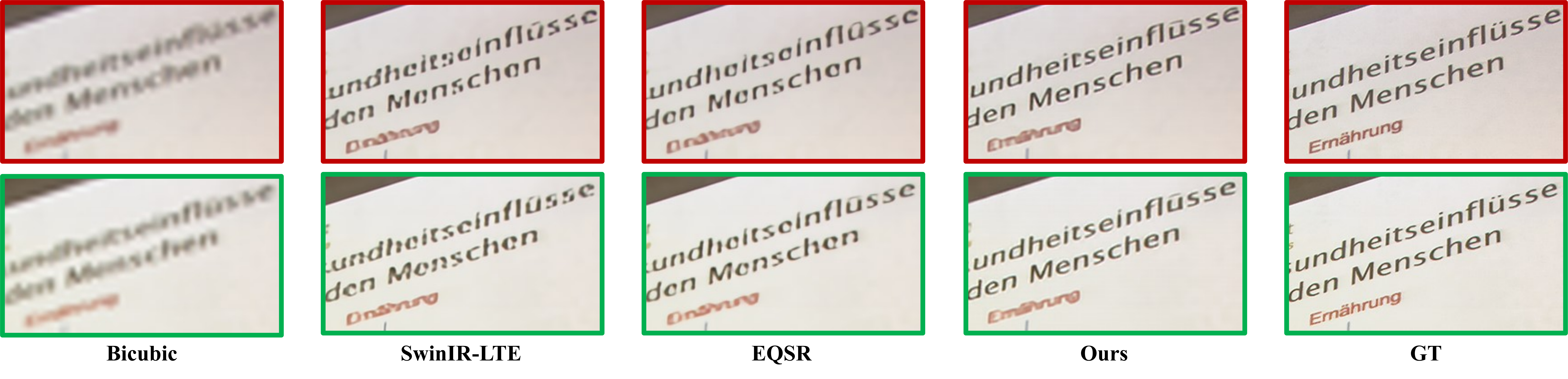}
\vspace{-3mm}
\caption{Visual results ($\times 6$) achieved by different methods on the Flickr1024~\cite{wang2019flickr1024} dataset. The images with red and green borders represent the left and right views respectively.}
% \vspace{-4mm}
\label{fig.comparison_with_SOTA2}
\end{figure*}

\section{Experiments}

\subsection{Implementation Details}
\noindent\textbf{Datasets}. Our training and validation datasets are provided by iPASSR, sourced from Middlebury~\cite{scharstein2014high} and Flickr1024~\cite{wang2019learning}. Specifically, we utilize 860 pairs of stereo images for training and 112 pairs for validation. To evaluate the generalization capability of our method, we employ the KITTI 2012~\cite{geiger2012we} and KITTI 2015~\cite{menze2015object} test sets for comprehensive assessment.

\noindent\textbf{Training Details}. We use $64\times 96$ patches as the inputs of the encoder. Let $B$ denote the training batch size, we first sample $B$ random scales $r^{i}$ following a continuous uniform distribution $\mathcal{U}(1,4)$, then we crop $B$ patches with sizes $\{64\times r^i, 96\times r^i\}^B_{i=1}$ from training image pairs. $64\times 96$ inputs are their down-sampled counterpart. For the ground truths, we converted these images to pixel samples (coordinate-RGB pairs) and we sample $64\times 96$ pixel samples for each of them so that the shapes of ground-truths are the same in a batch, as is done in \cite{chenLearningContinuousImage2021}. We use L1 loss and the Adam optimizer with an initial learning rate $5\times10^{-4}$. All models are trained for 200000 iterations with batch size $1$ per GPU, and the learning rate decays by the cosine annealing. In addition, we use SPyNet\cite{spynet} to estimate the disparity between the left and right views for spatial warping. SPyNet is usually used to estimate the optical flow of video sequences. Since the horizontal optical flow in stereo pairs is equivalent to the disparity under rectified settings, we use its horizontal optical flows to perform spatial warping. During training, we fine-tune only the stereo-aware adapter, scale-aware adapter, spatial adapter, and disparity-guided arbitrary scale upsampler, while keeping all other parameters frozen. 

\begin{table*}[!ht]
\centering
\caption{
Quantitative results achieved by different methods on the KITTI 2012~\cite{geiger2012we}, Middlebury~\cite{scharstein2014high}, and Flickr1024~\cite{wang2019learning} datasets. 
\# Tuned represents the number of trained parameters of the networks. 
Here, SCORE$\uparrow$, PSNR$\uparrow$, and SSIM$\uparrow$ values achieved on a pair of stereo images (i.e., (Left+Right)/2) are reported. 
\textcolor{red}{Red} and \textcolor{blue}{blue} colors indicate the best and the second-best performance, respectively.
}
\vspace{-2mm}
\resizebox{\textwidth}{!}{
    \begin{tabular}{lccccccccccc}
    \toprule
    \multirow{2}*{Method} & \multirow{2}*{Scale} & \multirow{2}*{\#Tuned} & \multicolumn{3}{c}{\textit{SCORE $\uparrow$}} & \multicolumn{3}{c}{\textit{PSNR $\uparrow$}} & \multicolumn{3}{c}{\textit{SSIM $\uparrow$}} \\
    \cmidrule(lr){4-6} \cmidrule(lr){7-9}\cmidrule(lr){10-12}
             &      &           & \makecell{KITTI2012} & \makecell{Middlebury} & Flickr1024 & \makecell{KITTI2012}& \makecell{Middlebury} & Flickr1024 & \makecell{KITTI2012}&  \makecell{Middlebury} & Flickr1024\\
    \midrule
    EDSR & $\times$2 & 38.6M & 0.8264& 0.8753& 0.7536 & 30.96& 34.95&28.66 & 0.9228 &0.9492& 0.9087 \\
    RDN & $\times$2 & 22.0M  & 0.8280 & 0.8757& 0.7488& 30.94& 34.94& 28.64& 0.9227 &0.9491&0.9084 \\
    StereoSR & $\times$2 &1.08M & 0.8061& 0.8543& 0.7247& 29.51&33.23&25.96&0.9073&0.9348&0.8599\\
    iPASSR & $\times$2 & 1.37M & 0.8266& 0.8663& 0.7365& 31.11&34.51&28.60&0.9240&0.9454&0.9097 \\
    SSRDE-FNet & $\times$2 & 2.10M & 0.8267& 0.8754& 0.7423&31.23&35.09&28.85&0.9254&0.9511& 0.9132\\
    NAFSSR-S & $\times$2 & 1.54M  & 0.8371&0.8812& 0.7747 & 31.38 & 35.30& 29.19 & 0.9266 &0.9514&0.9160	 \\
    SwinFIRSSR & $\times$2 & 23.94M  & \textcolor{red}{0.8502}& \textcolor{blue}{0.9022}& \textcolor{blue}{0.7751}& \textcolor{red}{31.79}& \textcolor{red}{36.52}& \textcolor{red}{30.14} & \textcolor{red}{0.9321}& \textcolor{red}{0.9598}& \textcolor{red}{0.9286}\\
    \midrule
    StereoINR (\textbf{Ours}) & $\times$2 & 2.03M  & \textcolor{blue}{0.8481}& \textcolor{red}{0.9025} & \textcolor{red}{0.7850} &\textcolor{blue}{31.58}& \textcolor{blue}{36.43}& \textcolor{blue}{30.02}&\textcolor{blue}{0.9289}&\textcolor{blue}{0.9590}&\textcolor{blue}{0.9254}\\
    \midrule
    \midrule
    EDSR &  $\times$4 & 38.9M &0.5724&0.6569&0.2860& 26.35&29.23&23.46&0.8015&0.8397&0.7285 \\
    RDN &  $\times$4 & 22.0M  &0.5716&0.6543&0.2815& 26.32&29.27&23.47&0.8014 &0.8404&0.7295\\
    StereoSR  &  $\times$4 & 1.42M   &0.5007&0.6021&0.2665& 24.53&27.64&21.70&0.7555&0.8022&0.6460 \\
    SRRes+SAM  &  $\times$4 & 1.73M  &0.5823&0.6455&\textcolor{blue}{0.3835}& 26.44&28.83&23.27&0.8018 &0.8290&0.7233 \\
    iPASSR  &  $\times$4 & 1.42M  & 0.5736&0.6474&0.3077& 26.56&29.16&23.44&0.8053&0.8367&0.7287\\
    SSRDE-FNet  & $\times$4 & 2.24M  & 0.5784&0.6500&0.3141&26.70&29.38&23.59&{0.8082}&0.8411&0.7352\\
    NAFSSR-S & $\times$4 & 1.56M  &0.5941&0.6665&0.3318& 26.93&29.72&23.88&0.8145&0.8490&0.7468 \\
    NAFSSR-L & $\times$4 & 23.79M  &0.6064&0.6893&0.3326& 27.12&30.20&24.17&0.8194&0.8605&0.7589 \\
    SwinFIRSSR & $\times$4 & 24.09M  & \textcolor{blue}{0.6116}&\textcolor{blue}{0.6981}&0.3211&\textcolor{red}{27.16} & \textcolor{blue}{30.44}&\textcolor{blue}{24.29} & \textcolor{red}{0.8235}& \textcolor{red}{0.8687}&\textcolor{red}{0.7681} \\
    \midrule
    StereoINR (\textbf{Ours}) & $\times$4 & 2.03M  & \textcolor{red}{0.6369}& \textcolor{red}{0.7145}& \textcolor{red}{0.5544}& \textcolor{blue}{27.14} &\textcolor{red}{30.47}& \textcolor{red}{24.32}& \textcolor{blue}{0.8214}&\textcolor{blue}{0.8674}&\textcolor{blue}{0.7674} \\
    \bottomrule
    \end{tabular}
}
\vspace{-2mm}
\label{tab:sota}
\end{table*}

\begin{figure*}[t]
\centering
\includegraphics[width=0.96  \textwidth]{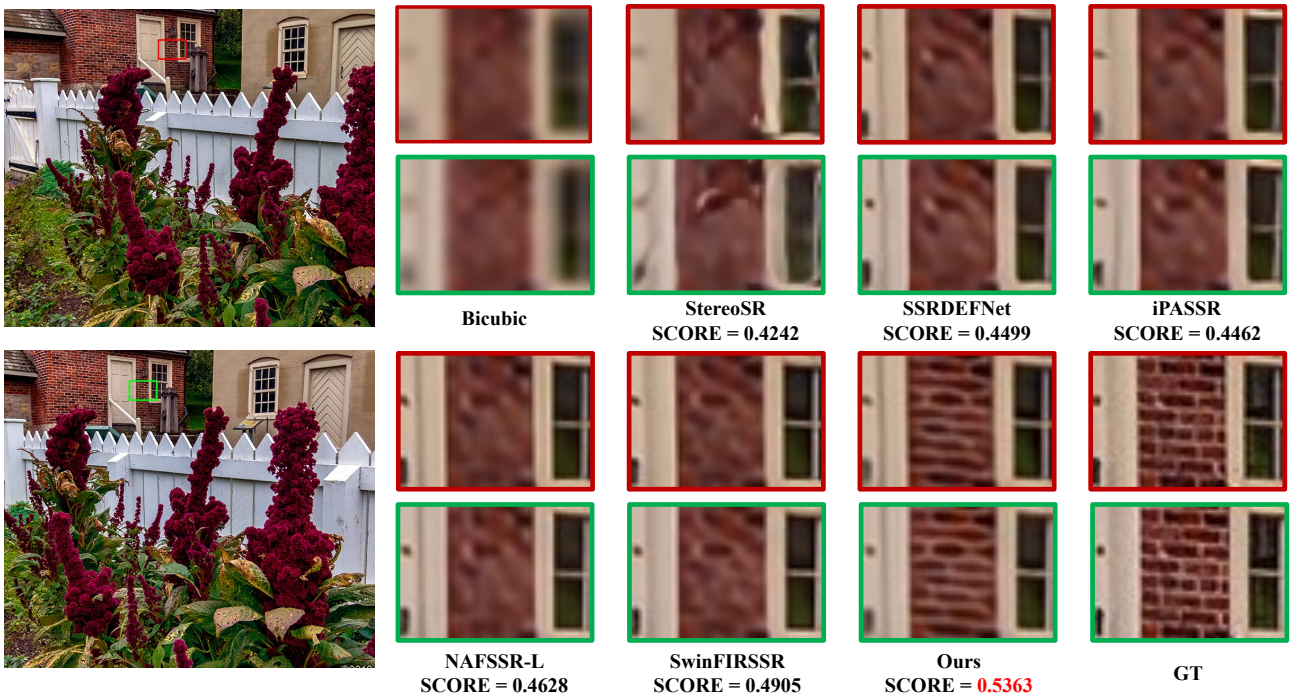}
\vspace{-4mm}
\caption{Visual results ($\times 4$) achieved by different methods on the Flickr 1024~\cite{wang2019learning} dataset. The images with red and green borders represent the left and right views respectively.}
\vspace{-2mm}
\label{fig.comparison_with_SOTA}
\end{figure*}

\begin{figure*}[t]
\centering
\includegraphics[width=0.96  \textwidth]{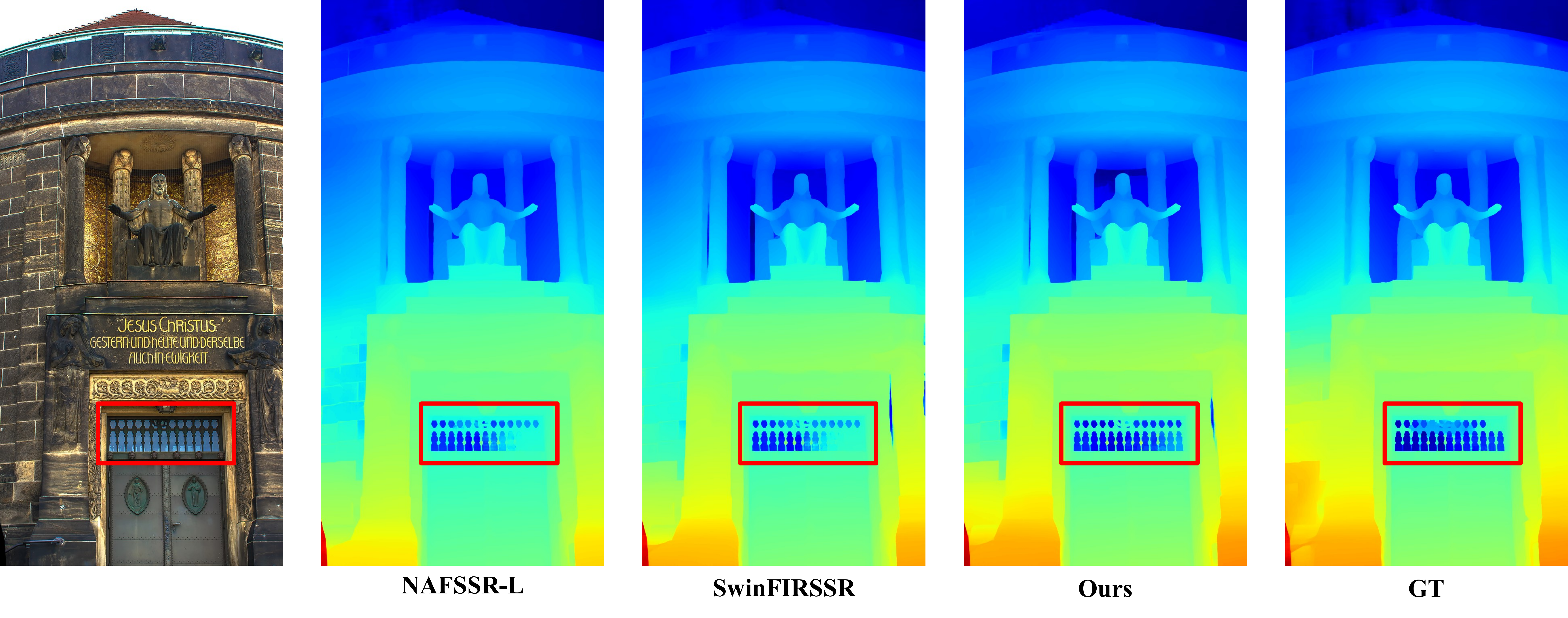}
\vspace{-4mm}
\caption{Visualization of disparity maps estimated from super-resolution results of different methods ($\times 4$) on the Flickr1024 dataset.}
\vspace{-4mm}
\label{fig.comparison_disp}
\end{figure*}

\noindent\textbf{Evaluation Metrics}. As the most widely used metrics for assessing image quality, PSNR and SSIM are not always consistent with the visual perception of the human eye. LPIPS~\cite{zhang2018unreasonable} is a learned perceptual image patch similarity metric designed to assess image quality in a manner that closely aligns with human visual perception. In order to measure the cross-view consistency of stereo images, NTIRE 2023 ~\cite{wang2023ntire} introduces Score to assess the quality of stereoscopic images by disparity:
\begin{align}
  SCORE =& 1-0.5\times(\mathcal{P}(I_L^{SR},I_L^{HR}) + \mathcal{P}(I_R^{SR},I_R^{HR})) \nonumber\\
  &-0.1\times \mathcal{L}(D^{SR},D^{HR}),
  \label{eq:eq_siqa}
\end{align}
Where $\mathcal{P}$ denotes LPIPS~\cite{zhang2018unreasonable} and $\mathcal{L}(D^{SR},D^{HR})$ denotes the MAE between disparity maps $D^{SR}$ and $D^{HR}$. To evaluate the stereo consistency, we used a pre-trained model such as RAFT-Stereo~\cite{teedRAFTRecurrentAllPairs2020} to estimate the disparity from SR and HR pairs. 

\subsection{Results}
Since INR provides a foundation for arbitrary-scale stereo super-resolution, we divide the quantitative experimental comparisons into two parts:
1) Out-scale: Comparison with other arbitrary-scale super-resolution methods beyond the training scales (×4+); 
2) In-scale: Comparison with current fixed-scale super-resolution methods on scales covered by the training data (×2–×4) to validate our method’s performance within trained scales.

\noindent \textbf{Quantitive results.} 
To validate the effectiveness of our method in both out-of-scale and in-scale stereo super-resolution, we conduct two sets of comparisons with state-of-the-art methods.

For out-of-scale upsampling, we compare our method with other INR-based methods, such as LIIF~\cite{chenLearningContinuousImage2021}, LTE~\cite{lee2022local}, LINF~\cite{yao2023local}, OPE-SR~\cite{songOPESROrthogonalPosition2023}, and EQSR~\cite{wang2023eqsr}, for arbitrary-scale upsampling, as shown in Table~\ref{tab:Quan_Middlebury}. For a fair comparison, we retrained the other methods on the stereo image dataset. Since large-scale super-resolution significantly degrades geometric information, comparing disparity results becomes less meaningful. Therefore, we evaluate performance using PSNR and perceptual loss (LPIPS), and our method achieves the best performance for out-of-scale upsampling.

For in-scale upsampling, we compare our method with fixed-scale super-resolution approaches, including EDSR~\cite{lim2017enhanced}, RDN~\cite{zhang2018residual}, StereoSR~\cite{jeon2018enhancing}, iPASSR~\cite{wang2021symmetric}, SSRDEFNet~\cite{dai2021feedback}, NAFSSR~\cite{chuNAFSSRStereoImage2022}, and SwinFIRSSR~\cite{zhangSwinFIRRevisitingSwinIR2023a}, as shown in Table~\ref{tab:sota}. Our method demonstrates competitive performance in terms of PSNR and SSIM while achieving a significant improvement in stereo consistency (SCORE). These results indicate that our approach successfully preserves multi-view geometric consistency during the upsampling process. Furthermore, although trained on the Flickr1024 and Middlebury datasets, our model also performs well on the KITTI dataset, indicating its generalization capability across different datasets.

\noindent \textbf{Qualitative results.} The in-scale qualitative results are shown in Figure \ref{fig.comparison_with_SOTA} and Figure \ref{fig.comparison_with_SOTA2}, where we compare our method with existing fixed-scale stereo super-resolution approaches at a $\times$4 scale. Our method effectively utilizes complementary information from different viewpoints to recover fine details, particularly in complex textures such as walls and grounds. We also annotate each image with its corresponding SCORE and observe that images with higher SCORE exhibit superior restoration of both textures and geometric structures, thereby validating the effectiveness of SCORE as a metric for stereo consistency.

Additionally, we evaluated the disparity results of super-resolution images generated by different methods. Using RAFT-Stereo~\cite{teedRAFTRecurrentAllPairs2020}, we estimate the disparity of the SR results, including those produced by SwinFIRSSR~\cite{zhangSwinFIRRevisitingSwinIR2023a} and NAFSSR~\cite{chuNAFSSRStereoImage2022}. As shown in Figure \ref{fig.comparison_disp}, our method produces more consistent cross-view geometry and generates more accurate disparity maps compared to other methods.

\begin{table}[!t]
\caption{Quantitative ablation study of \underline{different encoders} on Middlebury(PSNR $\uparrow$/SCORE $\uparrow$).}
\vspace{-2mm}
\resizebox{\linewidth}{!}
{
\centering
\setlength{\linewidth}{1.2pt}
\scriptsize
% \begin{tabular}{c|ccc|cc}
\begin{tabular}{c
|>{\centering\arraybackslash}p{0.30cm}>{\centering\arraybackslash}p{0.30cm}>{\centering\arraybackslash}p{0.30cm}
|>{\centering\arraybackslash}p{1.0cm}>{\centering\arraybackslash}p{1.0cm}
}
\toprule
Encoder & stereo & spatial & scale & $\times 2$ & $\times 4$ \\
\midrule\midrule
% flickr1024
\multirow{4}*{finetuned HAT} & \textcolor{gray}{\Checkmark} & \textcolor{gray}{\Checkmark} & \textcolor{gray}{\Checkmark} &  \textbf{36.43/0.9025} & \textbf{30.47/0.7145} \\
~ & \XSolidBrush & \textcolor{gray}{\Checkmark} & \textcolor{gray}{\Checkmark} & 35.91/0.8926  & 30.17/0.7020 \\
~ & \textcolor{gray}{\Checkmark} & \XSolidBrush & \textcolor{gray}{\Checkmark} & 36.01/0.9010  & 30.16/0.7091 \\
~ & \textcolor{gray}{\Checkmark} & \textcolor{gray}{\Checkmark} & \XSolidBrush & 36.14/0.9001  & 30.28/0.7144 \\
\midrule
vanilla HAT & \XSolidBrush & \XSolidBrush & \XSolidBrush & 35.90/0.8967  & 30.04/0.7103 \\
\bottomrule
\end{tabular}}
\label{tab:Quan_abl}
\end{table}

\subsection{Ablation Study}
\subsubsection{Latent encoder}
In this section, we take the vanilla HAT as the baseline encoder to investigate the impact of the stereo adapter, spatial adapter, and scale adapter. As demonstrated by the results in Table \ref{tab:Quan_abl}, the adapters offer significant performance improvements compared to the baseline. 
Compared to the vanilla HAT, the results indicate the importance of incorporating both cross-view information (introduced by the stereo adapter) and scale-aware information (extracted by the spatial adapter and scale adapter). 

\subsubsection{Disparity guided upsampler}
We propose that the disparity-guided upsampling method offers two key advantages. First, it explicitly captures the pixel-wise relationships between the left and right images, ensuring geometric consistency. Second, it introduces an inductive bias that enhances training efficiency and convergence by applying spatial warping to features.

To evaluate the performance of different upsampling methods, we replaced the upsampling module in StereoINR with pixel shuffle and retrained the model for comparison. As shown in the Table \ref{tab:Quan_upsampling}, the results demonstrate that our proposed DGASU method effectively improves the stereo consistency of the super-resolved results. 

% Demo ablation study
\begin{table}[!t]
\renewcommand{\arraystretch}{1.5}
\caption{Quantitative ablation study ($\times 4$) of StereoINR with \underline{\textbf{different upsampler}} on KITTI2012~\cite{geiger2012we}.}
\vspace{-2mm}
\resizebox{\linewidth}{!}
{
\centering
\setlength{\linewidth}{1.2pt}
\scriptsize
% \begin{tabular}{c|ccc|cc}
\begin{tabular}{c |>{\centering\arraybackslash}p{1.4cm}|>{\centering\arraybackslash}p{0.80cm}>{\centering\arraybackslash}p{0.80cm}
}
\toprule
 & Upsampler &  PSNR $\uparrow$ & SCORE $\uparrow$ \\
\midrule\midrule
% flickr1024
\multirow{2}*{StereoINR} &  DGASU & \textbf{24.32}  & \textbf{0.5544} \\
~  & pixel shuffle  & 24.25 &  0.4074 \\
\bottomrule
\end{tabular}}
\label{tab:Quan_upsampling}
\end{table}

\subsubsection{Other design}
In this section, We analyze the alternating attention in DGASU. As demonstrated in Table \ref{tab:Quan_attention}, leveraging cross-attention to interact between the left and right images significantly enhances cross-view consistency. Furthermore, using the Local Attribution Map (LAM) \cite{LAM}, we can visualize the importance of different pixels in the left and right views. As shown in Figure \ref{fig.lam_results}, cross-attention enables more right-view pixels to contribute to left-view reconstruction. We also calculate the corresponding Diffusion Index, and our results show that incorporating cross-attention allows the model to get higher DI and leverage a wider range of pixels in both views, demonstrating its effectiveness in enhancing feature interaction.

% Demo ablation study
\begin{table}[!t]
\renewcommand{\arraystretch}{1.5}
\caption{Quantitative ablation study of StereoINR with \underline{\textbf{different attention mechanism in DGASU}} on the Middlebury, KITTI2015, and Flickr1024 datasets (SCORE $\uparrow$).}
\vspace{-2mm}
\resizebox{\linewidth}{!}
{
\centering
\setlength{\linewidth}{1.2pt}
\scriptsize
% \begin{tabular}{c|ccc|cc}
\begin{tabular}{>{\centering\arraybackslash}p{1.3cm}|>{\centering\arraybackslash}p{1.0cm}>{\centering\arraybackslash}p{1.0cm}
>{\centering\arraybackslash}p{1.0cm}>{\centering\arraybackslash}p{1.0cm}
}
\toprule
 Methods & Middlebury & KITTI2015 & Flickr1024\\
\midrule\midrule
% flickr1024
 self+cross & \textbf{0.7145} &  \textbf{0.6939} & \textbf{0.5544} \\
 self+self  & 0.7057 & 0.6810  & 0.4579 \\
\bottomrule
\end{tabular}}
\label{tab:Quan_attention}
% \vspace{-15pt}
\end{table}

\begin{figure}[t]
\centering
\includegraphics[width=0.96  \linewidth]{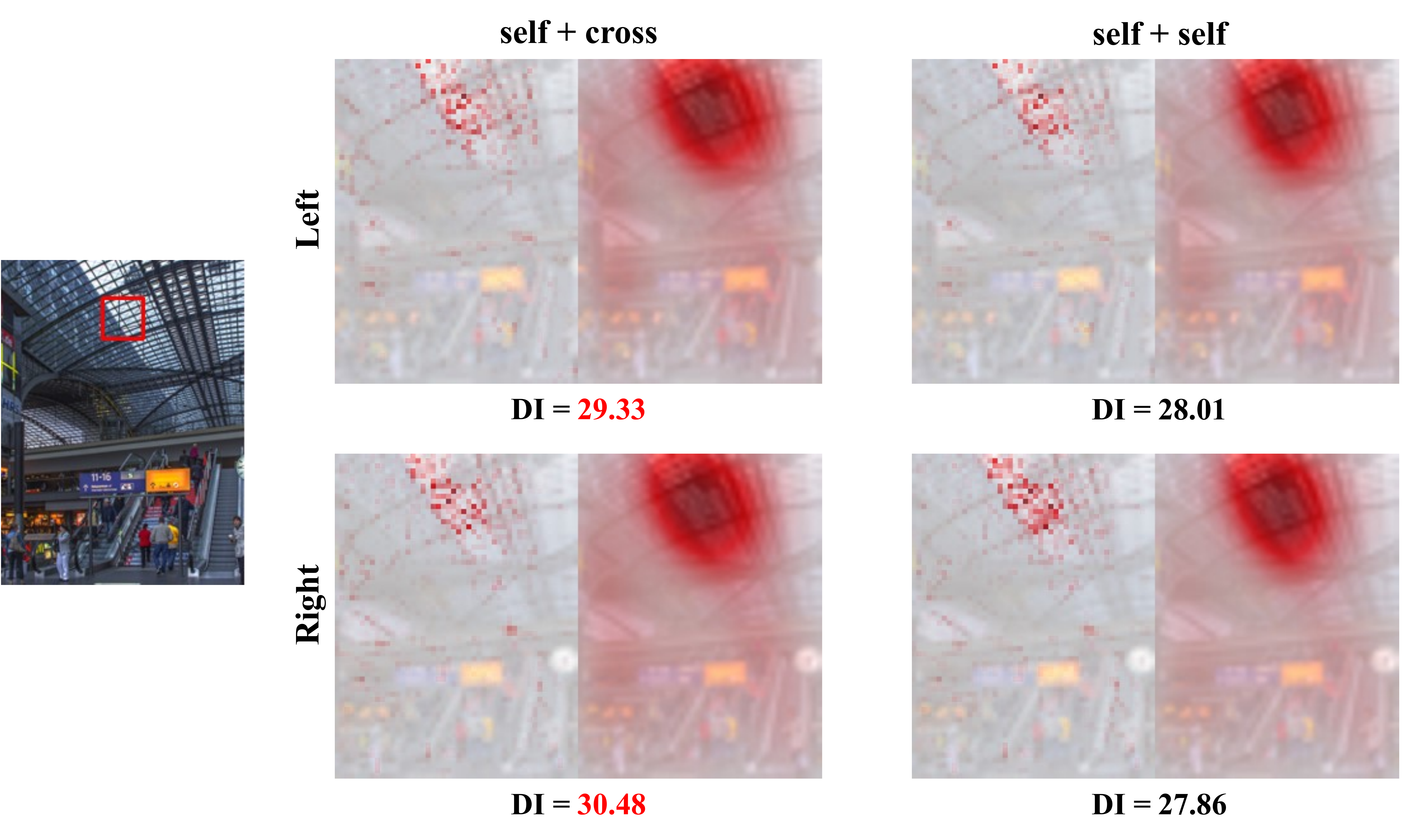}
\vspace{-2mm}
\caption{LAM analysis results of different methods on the Middlebury dataset. The first column shows the super-resolution results for the left view, while the second and third columns display the LAM results for different attentions, respectively. }
\label{fig.lam_results}
\end{figure}

\section{Conclusion}
In this paper, we propose StereoINR, a novel stereo super-resolution framework that enhances image quality while maintaining cross-view geometric consistency at arbitrary scales. Unlike traditional methods, StereoINR uses a disparity-guided upsampling strategy with implicit neural representation (INR) for flexible resolution enhancement. It incorporates stereo-aware, scale-aware, and spatial adapters into the Residual Hybrid Attention Group (RHAG) module for adaptive feature learning and an alternating self-attention and cross-attention mechanism to improve cross-view information interaction. Extensive experimental results demonstrate that our method significantly enhances cross-view consistency in stereo super-resolution and achieves arbitrary-scale reconstruction.

\noindent\textbf{Limitations and future work.}
Although the fine-tuning process involves relatively few parameters, the model's overall inference efficiency is still limited by the total number of parameters. Future work will explore more efficient approaches, such as lightweight architectures or knowledge distillation, to improve both performance and computational efficiency in stereo super-resolution models. Another limitation is that our method currently relies on accurate disparity estimation. Future work could explore self-supervised approach.

% \begin{acks}
% \end{acks}

\bibliographystyle{ACM-Reference-Format}
\bibliography{sample-sigconf}


\begin{thebibliography}{50}

%%% the Web address.

\ifx \showCODEN    \undefined \def \showCODEN     #1{\unskip}     \fi
\ifx \showDOI      \undefined \def \showDOI       #1{#1}\fi
\ifx \showISBNx    \undefined \def \showISBNx     #1{\unskip}     \fi
\ifx \showISBNxiii \undefined \def \showISBNxiii  #1{\unskip}     \fi
\ifx \showISSN     \undefined \def \showISSN      #1{\unskip}     \fi
\ifx \showLCCN     \undefined \def \showLCCN      #1{\unskip}     \fi
\ifx \shownote     \undefined \def \shownote      #1{#1}          \fi
\ifx \showarticletitle \undefined \def \showarticletitle #1{#1}   \fi
\ifx \showURL      \undefined \def \showURL       {\relax}        \fi
% invisible to TeX
\providecommand\bibfield[2]{#2}
\providecommand\bibinfo[2]{#2}
\providecommand\natexlab[1]{#1}
\providecommand\showeprint[2][]{arXiv:#2}

\bibitem[\protect\citeauthoryear{Cao, Wang, Xian, Li, Ni, Pi, Zhang, Zhang, Timofte, and Van~Gool}{Cao et~al\mbox{.}}{2023}]%
        {caoCiaoSRContinuousImplicit2023}
\bibfield{author}{\bibinfo{person}{Jiezhang Cao}, \bibinfo{person}{Qin Wang}, \bibinfo{person}{Yongqin Xian}, \bibinfo{person}{Yawei Li}, \bibinfo{person}{Bingbing Ni}, \bibinfo{person}{Zhiming Pi}, \bibinfo{person}{Kai Zhang}, \bibinfo{person}{Yulun Zhang}, \bibinfo{person}{Radu Timofte}, {and} \bibinfo{person}{Luc Van~Gool}.} \bibinfo{year}{2023}\natexlab{}.
\newblock \showarticletitle{Ciaosr: Continuous implicit attention-in-attention network for arbitrary-scale image super-resolution}. In \bibinfo{booktitle}{\emph{Proceedings of the IEEE/CVF Conference on Computer Vision and Pattern Recognition}}. \bibinfo{pages}{1796--1807}.
\newblock


\bibitem[\protect\citeauthoryear{Chen, Wang, Guo, Xu, Deng, Liu, Ma, Xu, Xu, and Gao}{Chen et~al\mbox{.}}{2021b}]%
        {chen2021pre}
\bibfield{author}{\bibinfo{person}{Hanting Chen}, \bibinfo{person}{Yunhe Wang}, \bibinfo{person}{Tianyu Guo}, \bibinfo{person}{Chang Xu}, \bibinfo{person}{Yiping Deng}, \bibinfo{person}{Zhenhua Liu}, \bibinfo{person}{Siwei Ma}, \bibinfo{person}{Chunjing Xu}, \bibinfo{person}{Chao Xu}, {and} \bibinfo{person}{Wen Gao}.} \bibinfo{year}{2021}\natexlab{b}.
\newblock \showarticletitle{Pre-trained image processing transformer}. In \bibinfo{booktitle}{\emph{Proceedings of the IEEE/CVF conference on computer vision and pattern recognition}}. \bibinfo{pages}{12299--12310}.
\newblock


\bibitem[\protect\citeauthoryear{Chen, Xu, Hong, Tsai, Kuo, and Lee}{Chen et~al\mbox{.}}{2023b}]%
        {chen2023cascaded}
\bibfield{author}{\bibinfo{person}{Hao-Wei Chen}, \bibinfo{person}{Yu-Syuan Xu}, \bibinfo{person}{Min-Fong Hong}, \bibinfo{person}{Yi-Min Tsai}, \bibinfo{person}{Hsien-Kai Kuo}, {and} \bibinfo{person}{Chun-Yi Lee}.} \bibinfo{year}{2023}\natexlab{b}.
\newblock \showarticletitle{Cascaded local implicit transformer for arbitrary-scale super-resolution}. In \bibinfo{booktitle}{\emph{Proceedings of the IEEE/CVF Conference on Computer Vision and Pattern Recognition}}. \bibinfo{pages}{18257--18267}.
\newblock


\bibitem[\protect\citeauthoryear{Chen, Pan, and Dong}{Chen et~al\mbox{.}}{2024}]%
        {chen2024bidirectional}
\bibfield{author}{\bibinfo{person}{Xiang Chen}, \bibinfo{person}{Jinshan Pan}, {and} \bibinfo{person}{Jiangxin Dong}.} \bibinfo{year}{2024}\natexlab{}.
\newblock \showarticletitle{Bidirectional multi-scale implicit neural representations for image deraining}. In \bibinfo{booktitle}{\emph{Proceedings of the IEEE/CVF Conference on Computer Vision and Pattern Recognition}}. \bibinfo{pages}{25627--25636}.
\newblock


\bibitem[\protect\citeauthoryear{Chen, Wang, Zhou, Qiao, and Dong}{Chen et~al\mbox{.}}{2023a}]%
        {chen2023activating}
\bibfield{author}{\bibinfo{person}{Xiangyu Chen}, \bibinfo{person}{Xintao Wang}, \bibinfo{person}{Jiantao Zhou}, \bibinfo{person}{Yu Qiao}, {and} \bibinfo{person}{Chao Dong}.} \bibinfo{year}{2023}\natexlab{a}.
\newblock \showarticletitle{Activating More Pixels in Image Super-Resolution Transformer}. In \bibinfo{booktitle}{\emph{Proceedings of the IEEE/CVF Conference on Computer Vision and Pattern Recognition (CVPR)}}. \bibinfo{pages}{22367--22377}.
\newblock


\bibitem[\protect\citeauthoryear{Chen, Liu, and Wang}{Chen et~al\mbox{.}}{2021a}]%
        {chenLearningContinuousImage2021}
\bibfield{author}{\bibinfo{person}{Yinbo Chen}, \bibinfo{person}{Sifei Liu}, {and} \bibinfo{person}{Xiaolong Wang}.} \bibinfo{year}{2021}\natexlab{a}.
\newblock \showarticletitle{Learning continuous image representation with local implicit image function}. In \bibinfo{booktitle}{\emph{Proceedings of the IEEE/CVF conference on computer vision and pattern recognition}}. \bibinfo{pages}{8628--8638}.
\newblock


\bibitem[\protect\citeauthoryear{Chen, Chen, Liu, Xu, Goel, Wang, Shi, and Wang}{Chen et~al\mbox{.}}{2022}]%
        {chenVideoINRLearningVideo2022}
\bibfield{author}{\bibinfo{person}{Zeyuan Chen}, \bibinfo{person}{Yinbo Chen}, \bibinfo{person}{Jingwen Liu}, \bibinfo{person}{Xingqian Xu}, \bibinfo{person}{Vidit Goel}, \bibinfo{person}{Zhangyang Wang}, \bibinfo{person}{Humphrey Shi}, {and} \bibinfo{person}{Xiaolong Wang}.} \bibinfo{year}{2022}\natexlab{}.
\newblock \showarticletitle{Videoinr: Learning video implicit neural representation for continuous space-time super-resolution}. In \bibinfo{booktitle}{\emph{Proceedings of the IEEE/CVF Conference on Computer Vision and Pattern Recognition}}. \bibinfo{pages}{2047--2057}.
\newblock


\bibitem[\protect\citeauthoryear{Chu, Chen, and Yu}{Chu et~al\mbox{.}}{2022}]%
        {chuNAFSSRStereoImage2022}
\bibfield{author}{\bibinfo{person}{Xiaojie Chu}, \bibinfo{person}{Liangyu Chen}, {and} \bibinfo{person}{Wenqing Yu}.} \bibinfo{year}{2022}\natexlab{}.
\newblock \showarticletitle{Nafssr: Stereo image super-resolution using nafnet}. In \bibinfo{booktitle}{\emph{Proceedings of the IEEE/CVF conference on computer vision and pattern recognition}}. \bibinfo{pages}{1239--1248}.
\newblock


\bibitem[\protect\citeauthoryear{Dai, Li, Yi, Fang, and Zhang}{Dai et~al\mbox{.}}{2021}]%
        {dai2021feedback}
\bibfield{author}{\bibinfo{person}{Qinyan Dai}, \bibinfo{person}{Juncheng Li}, \bibinfo{person}{Qiaosi Yi}, \bibinfo{person}{Faming Fang}, {and} \bibinfo{person}{Guixu Zhang}.} \bibinfo{year}{2021}\natexlab{}.
\newblock \showarticletitle{Feedback Network for Mutually Boosted Stereo Image Super-Resolution and Disparity Estimation}. In \bibinfo{booktitle}{\emph{Proceedings of the 29th ACM International Conference on Multimedia}}. \bibinfo{pages}{1985--1993}.
\newblock


\bibitem[\protect\citeauthoryear{Deng, Dong, Socher, Li, Li, and Fei-Fei}{Deng et~al\mbox{.}}{2009}]%
        {deng2009imagenet}
\bibfield{author}{\bibinfo{person}{Jia Deng}, \bibinfo{person}{Wei Dong}, \bibinfo{person}{Richard Socher}, \bibinfo{person}{Li-Jia Li}, \bibinfo{person}{Kai Li}, {and} \bibinfo{person}{Li Fei-Fei}.} \bibinfo{year}{2009}\natexlab{}.
\newblock \showarticletitle{Imagenet: A large-scale hierarchical image database}. In \bibinfo{booktitle}{\emph{2009 IEEE conference on computer vision and pattern recognition}}. Ieee, \bibinfo{pages}{248--255}.
\newblock


\bibitem[\protect\citeauthoryear{Geiger, Lenz, and Urtasun}{Geiger et~al\mbox{.}}{2012}]%
        {geiger2012we}
\bibfield{author}{\bibinfo{person}{Andreas Geiger}, \bibinfo{person}{Philip Lenz}, {and} \bibinfo{person}{Raquel Urtasun}.} \bibinfo{year}{2012}\natexlab{}.
\newblock \showarticletitle{Are we ready for autonomous driving? the kitti vision benchmark suite}. In \bibinfo{booktitle}{\emph{2012 IEEE conference on computer vision and pattern recognition}}. IEEE, \bibinfo{pages}{3354--3361}.
\newblock


\bibitem[\protect\citeauthoryear{Gong, Wang, Danelljan, Dai, and Van~Gool}{Gong et~al\mbox{.}}{2023}]%
        {gong2023continuous}
\bibfield{author}{\bibinfo{person}{Rui Gong}, \bibinfo{person}{Qin Wang}, \bibinfo{person}{Martin Danelljan}, \bibinfo{person}{Dengxin Dai}, {and} \bibinfo{person}{Luc Van~Gool}.} \bibinfo{year}{2023}\natexlab{}.
\newblock \showarticletitle{Continuous pseudo-label rectified domain adaptive semantic segmentation with implicit neural representations}. In \bibinfo{booktitle}{\emph{Proceedings of the IEEE/CVF Conference on Computer Vision and Pattern Recognition}}. \bibinfo{pages}{7225--7235}.
\newblock


\bibitem[\protect\citeauthoryear{Gu and Dong}{Gu and Dong}{2021}]%
        {LAM}
\bibfield{author}{\bibinfo{person}{Jinjin Gu} {and} \bibinfo{person}{Chao Dong}.} \bibinfo{year}{2021}\natexlab{}.
\newblock \showarticletitle{Interpreting Super-Resolution Networks with Local Attribution Maps}. In \bibinfo{booktitle}{\emph{2021 IEEE/CVF Conference on Computer Vision and Pattern Recognition (CVPR)}}. \bibinfo{pages}{9195--9204}.
\newblock


\bibitem[\protect\citeauthoryear{Jeon, Baek, Choi, and Kim}{Jeon et~al\mbox{.}}{2018}]%
        {jeon2018enhancing}
\bibfield{author}{\bibinfo{person}{Daniel~S Jeon}, \bibinfo{person}{Seung-Hwan Baek}, \bibinfo{person}{Inchang Choi}, {and} \bibinfo{person}{Min~H Kim}.} \bibinfo{year}{2018}\natexlab{}.
\newblock \showarticletitle{Enhancing the spatial resolution of stereo images using a parallax prior}. In \bibinfo{booktitle}{\emph{Proceedings of the IEEE conference on computer vision and pattern recognition}}. \bibinfo{pages}{1721--1730}.
\newblock


\bibitem[\protect\citeauthoryear{Jung, Hui, Luo, Yang, Liu, Yoo, Ranjan, and Demandolx}{Jung et~al\mbox{.}}{2023}]%
        {jungAnyFlowArbitraryScale2023}
\bibfield{author}{\bibinfo{person}{Hyunyoung Jung}, \bibinfo{person}{Zhuo Hui}, \bibinfo{person}{Lei Luo}, \bibinfo{person}{Haitao Yang}, \bibinfo{person}{Feng Liu}, \bibinfo{person}{Sungjoo Yoo}, \bibinfo{person}{Rakesh Ranjan}, {and} \bibinfo{person}{Denis Demandolx}.} \bibinfo{year}{2023}\natexlab{}.
\newblock \showarticletitle{Anyflow: Arbitrary scale optical flow with implicit neural representation}. In \bibinfo{booktitle}{\emph{Proceedings of the IEEE/CVF Conference on Computer Vision and Pattern Recognition}}. \bibinfo{pages}{5455--5465}.
\newblock


\bibitem[\protect\citeauthoryear{Lee and Jin}{Lee and Jin}{2022}]%
        {lee2022local}
\bibfield{author}{\bibinfo{person}{Jaewon Lee} {and} \bibinfo{person}{Kyong~Hwan Jin}.} \bibinfo{year}{2022}\natexlab{}.
\newblock \showarticletitle{Local texture estimator for implicit representation function}. In \bibinfo{booktitle}{\emph{Proceedings of the IEEE/CVF conference on computer vision and pattern recognition}}. \bibinfo{pages}{1929--1938}.
\newblock


\bibitem[\protect\citeauthoryear{Li, Lu, Qian, Lu, Zhang, and Jia}{Li et~al\mbox{.}}{2021}]%
        {li2021efficient}
\bibfield{author}{\bibinfo{person}{Wenbo Li}, \bibinfo{person}{Xin Lu}, \bibinfo{person}{Shengju Qian}, \bibinfo{person}{Jiangbo Lu}, \bibinfo{person}{Xiangyu Zhang}, {and} \bibinfo{person}{Jiaya Jia}.} \bibinfo{year}{2021}\natexlab{}.
\newblock \showarticletitle{On efficient transformer-based image pre-training for low-level vision}.
\newblock \bibinfo{journal}{\emph{arXiv preprint arXiv:2112.10175}} (\bibinfo{year}{2021}).
\newblock


\bibitem[\protect\citeauthoryear{Liang and Li}{Liang and Li}{2024}]%
        {liangAnyStereoArbitraryScale2024}
\bibfield{author}{\bibinfo{person}{Zhaohuai Liang} {and} \bibinfo{person}{Changhe Li}.} \bibinfo{year}{2024}\natexlab{}.
\newblock \showarticletitle{Any-Stereo: Arbitrary Scale Disparity Estimation for Iterative Stereo Matching}. In \bibinfo{booktitle}{\emph{Proceedings of the AAAI Conference on Artificial Intelligence}}, Vol.~\bibinfo{volume}{38}. \bibinfo{pages}{3333--3341}.
\newblock


\bibitem[\protect\citeauthoryear{Lim, Son, Kim, Nah, and Mu~Lee}{Lim et~al\mbox{.}}{2017}]%
        {lim2017enhanced}
\bibfield{author}{\bibinfo{person}{Bee Lim}, \bibinfo{person}{Sanghyun Son}, \bibinfo{person}{Heewon Kim}, \bibinfo{person}{Seungjun Nah}, {and} \bibinfo{person}{Kyoung Mu~Lee}.} \bibinfo{year}{2017}\natexlab{}.
\newblock \showarticletitle{Enhanced deep residual networks for single image super-resolution}. In \bibinfo{booktitle}{\emph{Proceedings of the IEEE conference on computer vision and pattern recognition workshops}}. \bibinfo{pages}{136--144}.
\newblock


\bibitem[\protect\citeauthoryear{Lugmayr, Danelljan, Van~Gool, and Timofte}{Lugmayr et~al\mbox{.}}{2020}]%
        {lugmayrSRFlowLearningSuperResolution2020}
\bibfield{author}{\bibinfo{person}{Andreas Lugmayr}, \bibinfo{person}{Martin Danelljan}, \bibinfo{person}{Luc Van~Gool}, {and} \bibinfo{person}{Radu Timofte}.} \bibinfo{year}{2020}\natexlab{}.
\newblock \showarticletitle{Srflow: Learning the super-resolution space with normalizing flow}. In \bibinfo{booktitle}{\emph{Proceedings of the European conference on computer vision (ECCV)}}. Springer, \bibinfo{pages}{715--732}.
\newblock


\bibitem[\protect\citeauthoryear{Ma, Yan, Tan, and Jiang}{Ma et~al\mbox{.}}{2021}]%
        {maPerceptionOrientedStereoImage2022}
\bibfield{author}{\bibinfo{person}{Chenxi Ma}, \bibinfo{person}{Bo Yan}, \bibinfo{person}{Weimin Tan}, {and} \bibinfo{person}{Xuhao Jiang}.} \bibinfo{year}{2021}\natexlab{}.
\newblock \showarticletitle{Perception-oriented stereo image super-resolution}. In \bibinfo{booktitle}{\emph{Proceedings of the 29th ACM International Conference on Multimedia}}. \bibinfo{pages}{2420--2428}.
\newblock


\bibitem[\protect\citeauthoryear{Menze and Geiger}{Menze and Geiger}{2015}]%
        {menze2015object}
\bibfield{author}{\bibinfo{person}{Moritz Menze} {and} \bibinfo{person}{Andreas Geiger}.} \bibinfo{year}{2015}\natexlab{}.
\newblock \showarticletitle{Object scene flow for autonomous vehicles}. In \bibinfo{booktitle}{\emph{Proceedings of the IEEE conference on computer vision and pattern recognition}}. \bibinfo{pages}{3061--3070}.
\newblock


\bibitem[\protect\citeauthoryear{Mildenhall, Srinivasan, Tancik, Barron, Ramamoorthi, and Ng}{Mildenhall et~al\mbox{.}}{2021}]%
        {mildenhall2021nerf}
\bibfield{author}{\bibinfo{person}{Ben Mildenhall}, \bibinfo{person}{Pratul~P Srinivasan}, \bibinfo{person}{Matthew Tancik}, \bibinfo{person}{Jonathan~T Barron}, \bibinfo{person}{Ravi Ramamoorthi}, {and} \bibinfo{person}{Ren Ng}.} \bibinfo{year}{2021}\natexlab{}.
\newblock \showarticletitle{Nerf: Representing scenes as neural radiance fields for view synthesis}.
\newblock \bibinfo{journal}{\emph{Commun. ACM}} \bibinfo{volume}{65}, \bibinfo{number}{1} (\bibinfo{year}{2021}), \bibinfo{pages}{99--106}.
\newblock


\bibitem[\protect\citeauthoryear{Qiu, He, Zhan, Pan, Xian, and Jin}{Qiu et~al\mbox{.}}{2023}]%
        {qiuSCNAFSSRPerceptualOrientedStereo2023}
\bibfield{author}{\bibinfo{person}{Zidian Qiu}, \bibinfo{person}{Zongyao He}, \bibinfo{person}{Zhihao Zhan}, \bibinfo{person}{Zilin Pan}, \bibinfo{person}{Xingyuan Xian}, {and} \bibinfo{person}{Zhi Jin}.} \bibinfo{year}{2023}\natexlab{}.
\newblock \showarticletitle{Sc-nafssr: Perceptual-oriented stereo image super-resolution using stereo consistency guided nafssr}. In \bibinfo{booktitle}{\emph{Proceedings of the IEEE/CVF Conference on Computer Vision and Pattern Recognition}}. \bibinfo{pages}{1426--1435}.
\newblock


\bibitem[\protect\citeauthoryear{Rahaman, Baratin, Arpit, Draxler, Lin, Hamprecht, Bengio, and Courville}{Rahaman et~al\mbox{.}}{2019}]%
        {rahaman2019spectral}
\bibfield{author}{\bibinfo{person}{Nasim Rahaman}, \bibinfo{person}{Aristide Baratin}, \bibinfo{person}{Devansh Arpit}, \bibinfo{person}{Felix Draxler}, \bibinfo{person}{Min Lin}, \bibinfo{person}{Fred Hamprecht}, \bibinfo{person}{Yoshua Bengio}, {and} \bibinfo{person}{Aaron Courville}.} \bibinfo{year}{2019}\natexlab{}.
\newblock \showarticletitle{On the spectral bias of neural networks}. In \bibinfo{booktitle}{\emph{International conference on machine learning}}. PMLR, \bibinfo{pages}{5301--5310}.
\newblock


\bibitem[\protect\citeauthoryear{Ranjan and Black}{Ranjan and Black}{2017}]%
        {spynet}
\bibfield{author}{\bibinfo{person}{Anurag Ranjan} {and} \bibinfo{person}{Michael~J. Black}.} \bibinfo{year}{2017}\natexlab{}.
\newblock \showarticletitle{Optical Flow Estimation Using a Spatial Pyramid Network}. In \bibinfo{booktitle}{\emph{2017 IEEE Conference on Computer Vision and Pattern Recognition (CVPR)}}. \bibinfo{pages}{2720--2729}.
\newblock


\bibitem[\protect\citeauthoryear{Sarkar, Hemani, Jain, Krishnamurthy, et~al\mbox{.}}{Sarkar et~al\mbox{.}}{2023}]%
        {sarkar2023parameter}
\bibfield{author}{\bibinfo{person}{Mausoom Sarkar}, \bibinfo{person}{Mayur Hemani}, \bibinfo{person}{Rishabh Jain}, \bibinfo{person}{Balaji Krishnamurthy}, {et~al\mbox{.}}} \bibinfo{year}{2023}\natexlab{}.
\newblock \showarticletitle{Parameter efficient local implicit image function network for face segmentation}. In \bibinfo{booktitle}{\emph{Proceedings of the IEEE/CVF Conference on Computer Vision and Pattern Recognition}}. \bibinfo{pages}{20970--20980}.
\newblock


\bibitem[\protect\citeauthoryear{Scharstein, Hirschm{\"u}ller, Kitajima, Krathwohl, Ne{\v{s}}i{\'c}, Wang, and Westling}{Scharstein et~al\mbox{.}}{2014}]%
        {scharstein2014high}
\bibfield{author}{\bibinfo{person}{Daniel Scharstein}, \bibinfo{person}{Heiko Hirschm{\"u}ller}, \bibinfo{person}{York Kitajima}, \bibinfo{person}{Greg Krathwohl}, \bibinfo{person}{Nera Ne{\v{s}}i{\'c}}, \bibinfo{person}{Xi Wang}, {and} \bibinfo{person}{Porter Westling}.} \bibinfo{year}{2014}\natexlab{}.
\newblock \showarticletitle{High-resolution stereo datasets with subpixel-accurate ground truth}. In \bibinfo{booktitle}{\emph{German conference on pattern recognition}}. Springer, \bibinfo{pages}{31--42}.
\newblock


\bibitem[\protect\citeauthoryear{Shang, Ren, Zhang, Fang, Zuo, and Ma}{Shang et~al\mbox{.}}{2024}]%
        {shang2024arbitrary}
\bibfield{author}{\bibinfo{person}{Wei Shang}, \bibinfo{person}{Dongwei Ren}, \bibinfo{person}{Wanying Zhang}, \bibinfo{person}{Yuming Fang}, \bibinfo{person}{Wangmeng Zuo}, {and} \bibinfo{person}{Kede Ma}.} \bibinfo{year}{2024}\natexlab{}.
\newblock \showarticletitle{Arbitrary-Scale Video Super-Resolution with Structural and Textural Priors}. In \bibinfo{booktitle}{\emph{European Conference on Computer Vision}}. Springer, \bibinfo{pages}{73--90}.
\newblock


\bibitem[\protect\citeauthoryear{Song, Sun, Zhang, Su, Shi, and He}{Song et~al\mbox{.}}{2023}]%
        {songOPESROrthogonalPosition2023}
\bibfield{author}{\bibinfo{person}{Gaochao Song}, \bibinfo{person}{Qian Sun}, \bibinfo{person}{Luo Zhang}, \bibinfo{person}{Ran Su}, \bibinfo{person}{Jianfeng Shi}, {and} \bibinfo{person}{Ying He}.} \bibinfo{year}{2023}\natexlab{}.
\newblock \showarticletitle{OPE-SR: Orthogonal position encoding for designing a parameter-free upsampling module in arbitrary-scale image super-resolution}. In \bibinfo{booktitle}{\emph{Proceedings of the IEEE/CVF Conference on Computer Vision and Pattern Recognition}}. \bibinfo{pages}{10009--10020}.
\newblock


\bibitem[\protect\citeauthoryear{Song, Choi, Jeong, and Sohn}{Song et~al\mbox{.}}{2020}]%
        {songStereoscopicImageSuperResolution2020}
\bibfield{author}{\bibinfo{person}{Wonil Song}, \bibinfo{person}{Sungil Choi}, \bibinfo{person}{Somi Jeong}, {and} \bibinfo{person}{Kwanghoon Sohn}.} \bibinfo{year}{2020}\natexlab{}.
\newblock \showarticletitle{Stereoscopic image super-resolution with stereo consistent feature}. In \bibinfo{booktitle}{\emph{Proceedings of the AAAI Conference on Artificial Intelligence}}, Vol.~\bibinfo{volume}{34}. \bibinfo{pages}{12031--12038}.
\newblock


\bibitem[\protect\citeauthoryear{Teed and Deng}{Teed and Deng}{2020}]%
        {teedRAFTRecurrentAllPairs2020}
\bibfield{author}{\bibinfo{person}{Zachary Teed} {and} \bibinfo{person}{Jia Deng}.} \bibinfo{year}{2020}\natexlab{}.
\newblock \showarticletitle{Raft: Recurrent all-pairs field transforms for optical flow}. In \bibinfo{booktitle}{\emph{Proceedings of the European conference on computer vision (ECCV)}}. Springer, \bibinfo{pages}{402--419}.
\newblock


\bibitem[\protect\citeauthoryear{Wang, Guo, Wang, Li, Gu, Timofte, Cheng, Ma, Ma, Sun, et~al\mbox{.}}{Wang et~al\mbox{.}}{2023b}]%
        {wang2023ntire}
\bibfield{author}{\bibinfo{person}{Longguang Wang}, \bibinfo{person}{Yulan Guo}, \bibinfo{person}{Yingqian Wang}, \bibinfo{person}{Juncheng Li}, \bibinfo{person}{Shuhang Gu}, \bibinfo{person}{Radu Timofte}, \bibinfo{person}{Ming Cheng}, \bibinfo{person}{Haoyu Ma}, \bibinfo{person}{Qiufang Ma}, \bibinfo{person}{Xiaopeng Sun}, {et~al\mbox{.}}} \bibinfo{year}{2023}\natexlab{b}.
\newblock \showarticletitle{NTIRE 2023 challenge on stereo image super-resolution: Methods and results}. In \bibinfo{booktitle}{\emph{Proceedings of the IEEE/CVF conference on computer vision and pattern recognition}}. \bibinfo{pages}{1346--1372}.
\newblock


\bibitem[\protect\citeauthoryear{Wang, Guo, Wang, Liang, Lin, Yang, and An}{Wang et~al\mbox{.}}{2020}]%
        {wangParallaxAttentionUnsupervised2022}
\bibfield{author}{\bibinfo{person}{Longguang Wang}, \bibinfo{person}{Yulan Guo}, \bibinfo{person}{Yingqian Wang}, \bibinfo{person}{Zhengfa Liang}, \bibinfo{person}{Zaiping Lin}, \bibinfo{person}{Jungang Yang}, {and} \bibinfo{person}{Wei An}.} \bibinfo{year}{2020}\natexlab{}.
\newblock \showarticletitle{Parallax attention for unsupervised stereo correspondence learning}.
\newblock \bibinfo{journal}{\emph{IEEE transactions on pattern analysis and machine intelligence}} \bibinfo{volume}{44}, \bibinfo{number}{4} (\bibinfo{year}{2020}), \bibinfo{pages}{2108--2125}.
\newblock


\bibitem[\protect\citeauthoryear{Wang, Wang, Liang, Lin, Yang, An, and Guo}{Wang et~al\mbox{.}}{2019a}]%
        {wang2019learning}
\bibfield{author}{\bibinfo{person}{Longguang Wang}, \bibinfo{person}{Yingqian Wang}, \bibinfo{person}{Zhengfa Liang}, \bibinfo{person}{Zaiping Lin}, \bibinfo{person}{Jungang Yang}, \bibinfo{person}{Wei An}, {and} \bibinfo{person}{Yulan Guo}.} \bibinfo{year}{2019}\natexlab{a}.
\newblock \showarticletitle{Learning parallax attention for stereo image super-resolution}. In \bibinfo{booktitle}{\emph{Proceedings of the IEEE/CVF Conference on Computer Vision and Pattern Recognition}}. \bibinfo{pages}{12250--12259}.
\newblock


\bibitem[\protect\citeauthoryear{Wang, Chen, Ni, Wang, Tong, and Liu}{Wang et~al\mbox{.}}{2023a}]%
        {wang2023eqsr}
\bibfield{author}{\bibinfo{person}{Xiaohang Wang}, \bibinfo{person}{Xuanhong Chen}, \bibinfo{person}{Bingbing Ni}, \bibinfo{person}{Hang Wang}, \bibinfo{person}{Zhengyan Tong}, {and} \bibinfo{person}{Yutian Liu}.} \bibinfo{year}{2023}\natexlab{a}.
\newblock \showarticletitle{Deep Arbitrary-Scale Image Super-Resolution via Scale-Equivariance Pursuit}. In \bibinfo{booktitle}{\emph{2023 IEEE/CVF Conference on Computer Vision and Pattern Recognition (CVPR)}}.
\newblock


\bibitem[\protect\citeauthoryear{Wang, Wang, Yang, An, and Guo}{Wang et~al\mbox{.}}{2019b}]%
        {wang2019flickr1024}
\bibfield{author}{\bibinfo{person}{Yingqian Wang}, \bibinfo{person}{Longguang Wang}, \bibinfo{person}{Jungang Yang}, \bibinfo{person}{Wei An}, {and} \bibinfo{person}{Yulan Guo}.} \bibinfo{year}{2019}\natexlab{b}.
\newblock \showarticletitle{Flickr1024: A large-scale dataset for stereo image super-resolution}. In \bibinfo{booktitle}{\emph{Proceedings of the IEEE/CVF International Conference on Computer Vision Workshops}}. \bibinfo{pages}{0--0}.
\newblock


\bibitem[\protect\citeauthoryear{Wang, Ying, Wang, Yang, An, and Guo}{Wang et~al\mbox{.}}{2021}]%
        {wang2021symmetric}
\bibfield{author}{\bibinfo{person}{Yingqian Wang}, \bibinfo{person}{Xinyi Ying}, \bibinfo{person}{Longguang Wang}, \bibinfo{person}{Jungang Yang}, \bibinfo{person}{Wei An}, {and} \bibinfo{person}{Yulan Guo}.} \bibinfo{year}{2021}\natexlab{}.
\newblock \showarticletitle{Symmetric parallax attention for stereo image super-resolution}. In \bibinfo{booktitle}{\emph{Proceedings of the IEEE/CVF Conference on Computer Vision and Pattern Recognition}}. \bibinfo{pages}{766--775}.
\newblock


\bibitem[\protect\citeauthoryear{Wei and Zhang}{Wei and Zhang}{2023}]%
        {Wei_2023_CVPR}
\bibfield{author}{\bibinfo{person}{Min Wei} {and} \bibinfo{person}{Xuesong Zhang}.} \bibinfo{year}{2023}\natexlab{}.
\newblock \showarticletitle{Super-Resolution Neural Operator}. In \bibinfo{booktitle}{\emph{Proceedings of the IEEE/CVF Conference on Computer Vision and Pattern Recognition (CVPR)}}. \bibinfo{pages}{18247--18256}.
\newblock


\bibitem[\protect\citeauthoryear{Xu, Wang, and Shi}{Xu et~al\mbox{.}}{2021}]%
        {xuUltraSRSpatialEncoding2022}
\bibfield{author}{\bibinfo{person}{Xingqian Xu}, \bibinfo{person}{Zhangyang Wang}, {and} \bibinfo{person}{Humphrey Shi}.} \bibinfo{year}{2021}\natexlab{}.
\newblock \showarticletitle{Ultrasr: Spatial encoding is a missing key for implicit image function-based arbitrary-scale super-resolution}.
\newblock \bibinfo{journal}{\emph{arXiv preprint arXiv:2103.12716}} (\bibinfo{year}{2021}).
\newblock


\bibitem[\protect\citeauthoryear{Yang, Shen, Yue, and Li}{Yang et~al\mbox{.}}{2021}]%
        {yangImplicitTransformerNetwork}
\bibfield{author}{\bibinfo{person}{Jingyu Yang}, \bibinfo{person}{Sheng Shen}, \bibinfo{person}{Huanjing Yue}, {and} \bibinfo{person}{Kun Li}.} \bibinfo{year}{2021}\natexlab{}.
\newblock \showarticletitle{Implicit transformer network for screen content image continuous super-resolution}.
\newblock \bibinfo{journal}{\emph{Advances in Neural Information Processing Systems}}  \bibinfo{volume}{34} (\bibinfo{year}{2021}), \bibinfo{pages}{13304--13315}.
\newblock


\bibitem[\protect\citeauthoryear{Yao, Hong, Chanussot, Meng, Zhu, and Xu}{Yao et~al\mbox{.}}{2020}]%
        {yaoCrossAttentionCoupledUnmixing2020}
\bibfield{author}{\bibinfo{person}{Jing Yao}, \bibinfo{person}{Danfeng Hong}, \bibinfo{person}{Jocelyn Chanussot}, \bibinfo{person}{Deyu Meng}, \bibinfo{person}{Xiaoxiang Zhu}, {and} \bibinfo{person}{Zongben Xu}.} \bibinfo{year}{2020}\natexlab{}.
\newblock \showarticletitle{Cross-attention in coupled unmixing nets for unsupervised hyperspectral super-resolution}. In \bibinfo{booktitle}{\emph{Proceedings of the European conference on computer vision (ECCV)}}. Springer, \bibinfo{pages}{208--224}.
\newblock


\bibitem[\protect\citeauthoryear{Yao, Tsao, Lo, Tseng, Chang, and Lee}{Yao et~al\mbox{.}}{2023}]%
        {yao2023local}
\bibfield{author}{\bibinfo{person}{Jie-En Yao}, \bibinfo{person}{Li-Yuan Tsao}, \bibinfo{person}{Yi-Chen Lo}, \bibinfo{person}{Roy Tseng}, \bibinfo{person}{Chia-Che Chang}, {and} \bibinfo{person}{Chun-Yi Lee}.} \bibinfo{year}{2023}\natexlab{}.
\newblock \showarticletitle{Local Implicit Normalizing Flow for Arbitrary-Scale Image Super-Resolution}. In \bibinfo{booktitle}{\emph{Proceedings of the IEEE/CVF Conference on Computer Vision and Pattern Recognition}}.
\newblock


\bibitem[\protect\citeauthoryear{Ying, Wang, Wang, Sheng, An, and Guo}{Ying et~al\mbox{.}}{2020}]%
        {ying2020stereo}
\bibfield{author}{\bibinfo{person}{Xinyi Ying}, \bibinfo{person}{Yingqian Wang}, \bibinfo{person}{Longguang Wang}, \bibinfo{person}{Weidong Sheng}, \bibinfo{person}{Wei An}, {and} \bibinfo{person}{Yulan Guo}.} \bibinfo{year}{2020}\natexlab{}.
\newblock \showarticletitle{A stereo attention module for stereo image super-resolution}.
\newblock \bibinfo{journal}{\emph{IEEE Signal Processing Letters}}  \bibinfo{volume}{27} (\bibinfo{year}{2020}), \bibinfo{pages}{496--500}.
\newblock


\bibitem[\protect\citeauthoryear{Zhang, Huang, Liu, Wang, and Jin}{Zhang et~al\mbox{.}}{2022}]%
        {zhangSwinFIRRevisitingSwinIR2023a}
\bibfield{author}{\bibinfo{person}{Dafeng Zhang}, \bibinfo{person}{Feiyu Huang}, \bibinfo{person}{Shizhuo Liu}, \bibinfo{person}{Xiaobing Wang}, {and} \bibinfo{person}{Zhezhu Jin}.} \bibinfo{year}{2022}\natexlab{}.
\newblock \showarticletitle{Swinfir: Revisiting the swinir with fast fourier convolution and improved training for image super-resolution}.
\newblock \bibinfo{journal}{\emph{arXiv preprint arXiv:2208.11247}} (\bibinfo{year}{2022}).
\newblock


\bibitem[\protect\citeauthoryear{Zhang, Isola, Efros, Shechtman, and Wang}{Zhang et~al\mbox{.}}{2018a}]%
        {zhang2018unreasonable}
\bibfield{author}{\bibinfo{person}{Richard Zhang}, \bibinfo{person}{Phillip Isola}, \bibinfo{person}{Alexei~A Efros}, \bibinfo{person}{Eli Shechtman}, {and} \bibinfo{person}{Oliver Wang}.} \bibinfo{year}{2018}\natexlab{a}.
\newblock \showarticletitle{The unreasonable effectiveness of deep features as a perceptual metric}. In \bibinfo{booktitle}{\emph{Proceedings of the IEEE conference on computer vision and pattern recognition}}. \bibinfo{pages}{586--595}.
\newblock


\bibitem[\protect\citeauthoryear{Zhang, Yu, Jiang, Nie, Yao, Huang, and Tao}{Zhang et~al\mbox{.}}{2024}]%
        {zhangStereoImageRestoration2024}
\bibfield{author}{\bibinfo{person}{Shengping Zhang}, \bibinfo{person}{Wei Yu}, \bibinfo{person}{Feng Jiang}, \bibinfo{person}{Liqiang Nie}, \bibinfo{person}{Hongxun Yao}, \bibinfo{person}{Qingming Huang}, {and} \bibinfo{person}{Dacheng Tao}.} \bibinfo{year}{2024}\natexlab{}.
\newblock \showarticletitle{Stereo image restoration via attention-guided correspondence learning}.
\newblock \bibinfo{journal}{\emph{IEEE Transactions on Pattern Analysis and Machine Intelligence}} (\bibinfo{year}{2024}).
\newblock


\bibitem[\protect\citeauthoryear{Zhang, Tian, Kong, Zhong, and Fu}{Zhang et~al\mbox{.}}{2018b}]%
        {zhang2018residual}
\bibfield{author}{\bibinfo{person}{Yulun Zhang}, \bibinfo{person}{Yapeng Tian}, \bibinfo{person}{Yu Kong}, \bibinfo{person}{Bineng Zhong}, {and} \bibinfo{person}{Yun Fu}.} \bibinfo{year}{2018}\natexlab{b}.
\newblock \showarticletitle{Residual dense network for image super-resolution}. In \bibinfo{booktitle}{\emph{Proceedings of the IEEE conference on computer vision and pattern recognition}}. \bibinfo{pages}{2472--2481}.
\newblock


\bibitem[\protect\citeauthoryear{Zhou, Xue, Deng, Zhang, Gao, and Tong}{Zhou et~al\mbox{.}}{2024}]%
        {zhou2024asteisr}
\bibfield{author}{\bibinfo{person}{Yuanbo Zhou}, \bibinfo{person}{Yuyang Xue}, \bibinfo{person}{Wei Deng}, \bibinfo{person}{Xinlin Zhang}, \bibinfo{person}{Qinquan Gao}, {and} \bibinfo{person}{Tong Tong}.} \bibinfo{year}{2024}\natexlab{}.
\newblock \showarticletitle{ASteISR: Adapting Single Image Super-resolution Pre-trained Model for Efficient Stereo Image Super-resolution}.
\newblock \bibinfo{journal}{\emph{arXiv preprint arXiv:2407.03598}} (\bibinfo{year}{2024}).
\newblock


\bibitem[\protect\citeauthoryear{Zou, Gao, Chen, Zhang, Jiang, Yu, and Tan}{Zou et~al\mbox{.}}{2023}]%
        {zouCrossViewHierarchyNetwork2023}
\bibfield{author}{\bibinfo{person}{Wenbin Zou}, \bibinfo{person}{Hongxia Gao}, \bibinfo{person}{Liang Chen}, \bibinfo{person}{Yunchen Zhang}, \bibinfo{person}{Mingchao Jiang}, \bibinfo{person}{Zhongxin Yu}, {and} \bibinfo{person}{Ming Tan}.} \bibinfo{year}{2023}\natexlab{}.
\newblock \showarticletitle{Cross-view hierarchy network for stereo image super-resolution}. In \bibinfo{booktitle}{\emph{Proceedings of the IEEE/CVF Conference on Computer Vision and Pattern Recognition}}. \bibinfo{pages}{1396--1405}.
\newblock


\end{thebibliography}

\end{document}